\documentclass[11pt]{article}
\usepackage[utf8]{inputenc}
\usepackage[english]{babel}
\usepackage{amssymb}
\usepackage{mathrsfs}
\usepackage{stmaryrd}
\usepackage{tabularx}
\usepackage{booktabs}
\usepackage{times}
\usepackage{setspace}
\usepackage{authblk}
\usepackage{placeins}
\usepackage{xfrac}
\usepackage{graphicx}
\usepackage{float}
\usepackage[T1]{fontenc}
\usepackage[final]{pdfpages}
\usepackage{pdflscape}
\usepackage[a4paper,top=3cm,bottom=2cm,left=3cm,right=3cm,marginparwidth=1.75cm]{geometry}
\usepackage{amsmath, amssymb}
\usepackage{amsthm}
\usepackage[colorinlistoftodos]{todonotes}
\usepackage[colorlinks=true, allcolors=black]{hyperref}

\usepackage[tight]{shorttoc}

\usepackage{hyperref}
\usepackage{mathabx}
\usepackage[bottom]{footmisc}
\usepackage{eurosym}
\usepackage{natbib}

\usepackage{tocloft}
\newcommand{\R}{\mathbb{R}}

\newcommand{\C}{\mathbb{C}}
\newcommand{\N}{\mathbb{N}}
\newcommand{\E}{\mathbb{E}}
\newcommand{\V}{\mathbb{V}}
\newcommand{\p}{\mathbb{P}}

\newcommand{\cov}{\operatorname{Cov}}

\newcommand{\cP}{{\mathcal P}}

\newcommand{\e}{\varepsilon}

\newcommand{\cB}{{\mathcal B}}

\newcommand{\cM}{{\mathcal M}}

\newcommand{\cC}{{\mathcal C}}

\newcommand{\cL}{{\mathcal L}}
\newcommand{\cR}{{\mathcal R}}

\newcommand{\pS}{{\psi_{sum}}}
\newcommand{\pSn}{{\psi_{sum}^T}}
\newcommand{\pSC}{{\psi_{sum}^C}}
\newcommand{\pO}{{\psi_{or}}}
\newcommand{\pOn}{{\psi_{or}^T}}
\newcommand{\pOC}{{\psi_{or}^C}}
\newcommand{\pA}{{\psi_{and}}}
\newcommand{\pAn}{{\psi_{and}^T}}
\newcommand{\pAC}{{\psi_{and}^C}}

\everymath{\displaystyle}

\theoremstyle{definition}

\newtheorem{defi}{Definition}[section]
\newtheorem{lem}{Lemma}[section]
\newtheorem{prop}{Proposition}[section]
\newtheorem{cor}{Corollary}[section]
\newtheorem{rem}{Remark}[section]

\usepackage{hyperref}
\hypersetup{	
colorlinks=true, 
breaklinks=true, 
urlcolor= purple, 
linkcolor= black,	
citecolor= brown,	
}

\usepackage{lipsum}

\title{Joint modelling of financial cost and carbon footprint in motor insurance claims: a BINAR(1) decision framework for circular auto parts adoption}

\author[1,2]{Charles Minier\thanks{charles.minier41@gmail.com}}
\author[3]{Stéphane Loisel\thanks{stephane.loisel@lecnam.net}}
\author[4,5,6]{Renata Alcoforado\thanks{alcoforado.renata@ufpe.br}}

\affil[1]{Université Claude Bernard Lyon 1, Laboratoire de Sciences Actuarielle et Financière, Institut de Science Financière et d’Assurances (ISFA), Lyon, France}
\affil[2]{Chaire DIALog, Fondation du Risque, Institut Louis Bachelier, Paris}
\affil[3]{Conservatoire National des Arts et Métiers (CNAM), France}
\affil[4]{Universidade Federal de Pernambuco, Brazil}
\affil[5]{Chaire d’Excellence ACTIONS \& Institut de Mathématiques de Marseille - Aix Marseille Université, France}
\affil[6]{CEMAPRE \& ISEG, Universidade de Lisboa, Portugal}

\begin{document}

\maketitle

\begin{abstract}
Motor insurance claims management increasingly faces a dual objective: controlling claim costs while reducing greenhouse-gas emissions associated with vehicle repair decisions. This paper proposes a bivariate, discrete-time modelling framework that links the dynamics of financial claims and repair-related carbon footprint 
inspired by a BINAR(1) dependence structure, allowing for temporal persistence, cross-dependence and a positive/negative correlation between the two dimensions.
Using externally sourced emission factors and cost information for replacement parts, we operationalise the carbon footprint of claims under two repair pathways: new parts versus circular-economy reused parts (PIEC). We then conduct scenario analyses in which the acceptance rate of reused parts varies, and we quantify the joint implications for expected monetary outlays and CO$_2$-equivalent emissions. Results show that higher PIEC acceptance can simultaneously reduce expected emissions and average claim cost, and the magnitude of the joint benefit depends on the dependence structure and persistence captured by the BINAR(1) process. The proposed framework provides a transparent, reproducible tool for insurers to evaluate repair-policy levers and to align claims operations with decarbonisation targets while retaining an explicit stochastic risk perspective.

\smallskip

\end{abstract}

\vspace{1em}

\noindent \textbf{Keywords:} BINAR(1); Bivariate Risk Model; Carbon Footprint; Sustainable Insurance; Circular Economy; Claim Management.

\newpage
\tableofcontents

\newpage
\section{Introduction}

The growing awareness of environmental concerns has recently brought attention to the carbon footprint associated with claim management in motor insurance. Notably, French insurers and industry bodies have addressed this issue through publications such as the \textit{Livre Blanc} (\cite{livreblanc}), and companies like Allianz have contributed with detailed assessments of operational emissions (\cite{Allianzrapport2023}). In France, where the transport sector remains the leading source of greenhouse gas emissions, accounting for a substantial portion of national totals, reducing environmental impact in this domain is both urgent and strategic. 

In June 2012, the United Nations set the Principles of Sustainable Insurance, which highlighted sustainability as a key issue in actuarial science since then. In relation to these principles, the Partnership for Carbon Accounting Financials (PCAF) has notably developed standards to measure insurance-associated emissions, delivered in 2022 as the "Global GHG Accounting and Reporting Standard for Insurance-Associated Emissions", completed in the document \cite{PCAF} with detailed methodological guidance for the measurement and disclosure.
In addition, the report \cite{CrawfordCompany} discusses the possibilities of using green materials or processes to resolve claims, by emphasising the difficulties when sustainable methods are more expensive than "like-for-like" repairs. 
As for more particular studies, for example,
\cite{stricker2022green} develops a comprehensive roadmap along the insurance value chain for executive management to design the sustainability efforts of a company, with special focus on property and casualty, and \cite{dawson2022assessing} aimed at evaluating the emissions and reduction of greenhouse gas by the whole activity of Allianz at the end of the 2010s.

Dwelling specifically on the emissions coming from the life cycle of a car and going further than a pure insurance context, an independent label,
the Eco Repair Score, has been designed by a Belgian society to promote informed ecological choices, allowing consumers to assess easily online the environmental impact of one's repair, \cite{EcoRepairScore}.
Numerous related studies can be found in the scientific literature: for instance, \cite{kurzmann2025emission} examines emission reduction dynamics within the vehicle repair industry by distinguishing between traditional or cleaner solutions, 
\cite{chen2020environmental} analyses the environmental impact of ignition coils and tries to find out how to diminish the related carbon emissions, while
\cite{jovanovic2023driving} provides a review of the technological development of electric or autonomous vehicles, with special attention on carbon footprints and the implementation of additive manufacturing using recycled materials.

Furthermore, since 2024, the European Corporate Sustainability Reporting Directive (CSRD) mandates certain corporations to disclose carbon emissions. This has increased the relevance of integrating non-financial risks, such as environmental impact, into traditional financial risk models. In this context, the relationship between the monetary cost of managing motor claims—especially those involving collisions—and their corresponding carbon emissions emerges as a key issue. The theoretical underpinnings of non-financial risk and its intersection with insurance are explored in works such as \cite{rey2003note}, \cite{eeckhoudt2007good}, and \cite{courbage2007precautionary}. Moreover, \cite{gong2024examining} explored recently potential strategies developed by insurance companies for addressing carbon emissions in the automotive supply chain in the United Kingdom.

Motivated by these developments, this paper presents a bivariate risk modelling approach that jointly captures financial losses and carbon emissions. Our objective is to represent the dual risk faced by insurers: monetary costs and environmental liabilities. 

While multivariate models have long been used to account for dependencies across insurance lines (see, for example, \cite{biard2013asymptotic}, \cite{cai2007dependence}, and \cite{konstantinides2016asymptotic}), these studies typically represent either common shock structures or claim amount correlations. Bidimensional risk models, such as those in \cite{avram2008exit} and \cite{burnecki2021ruin}, often consider insurer-reinsurer dynamics or claim splitting. Some discrete-time frameworks, such as \cite{ma2015bidimensional}, introduce common shocks through autoregressive time series, but most of these contributions assume only one type of claim per line, which limits their application to more complex operational realities.

To overcome this, we draw inspiration from refinements to the classical Cramér-Lundberg model that introduce secondary claims or “by-claims”, i.e. losses that follow an initial main claim. This reflects real-world patterns in motor insurance, where a single incident can generate both immediate material damage claims and delayed bodily injury claims. Models such as \cite{waters1985ruin}, \cite{yuen2001ruin}, and \cite{yuen2005ultimate} incorporate by-claims either as delayed responses or as secondary claims linked to the initial one. More recent extensions \cite{macci2006large, fu2016asymptotic, yang2019asymptotic} build on these ideas, but typically assume a fixed number of by-claims per main claim or independence between them - assumptions that may not hold in real-world operations.

Furthermore, discrete-time models based on thinning operators, such as INAR or INMA processes, allow for richer stochastic dynamics and claim clustering. These have been employed in a variety of univariate contexts, in \cite{Cossette2010DiscreteTimeRM, fu2016asymptotic, hu2018risk, guan2022analysis}, or in \cite{minier2026risk}, with the last study capturing random by-claims using autoregressive time series. However, extending these insights to a multidimensional setting--especially one that jointly models financial and environmental risk--remains underexplored.

To address this gap, we introduce two discrete-time bivariate risk models. The first extends the framework from \cite{minier2026risk} by using a bivariate integer-valued autoregressive process of order 1 (BINAR(1)), and is a natural, general extension of existing models.
The second model builds on the same intuition but adapts its structure to better reflect the specific features of carbon emissions in motor claim operations, taking into account the claim-by-claim nature of emissions and costs, in a more particular way: 
it considers jointly both financial and ecological risks, with the possibility of subsequent losses through by-claims.

We examine both models in parallel, comparing their probabilistic properties and asymptotic behaviour. We derive expressions for the covariance structures related to the time series, the net profit conditions, and some ruin probabilities. The exploration as a ruin framework enables to consider not only the financial solvency limits but also potential excess in terms of carbon emission.
In cases where claim size distributions are light-tailed, we provide the corresponding Lundberg equations and adjustment coefficients.
Finally, using data from Allianz and the \textit{Livre Blanc}, we simulate scenarios involving the substitution of new car parts with recycled alternatives (PIEC). These simulations offer insight into the trade-offs between financial costs and carbon emissions in post-collision claim scenarios.
Our study shows that the properties of the first process can be used as a benchmark for more sophisticated and more specific models like the second one, which could be studied and calibrated thanks to real data when it becomes available, and also because some proofs share similar steps for both types of processes. 

This study contributes to carbon-footprint management in service operations in three ways. First, it introduces a stochastic bivariate framework that jointly represents financial and repair-related carbon outcomes, enabling decision evaluation under dependence and temporal persistence rather than via independent or static emission accounting, while also exploring the model's properties, such as ruin probabilities, covariances and asymptotic analyses. Second, it translates part-level replacement choices into an operational carbon-footprint metric that can be stress-tested under realistic policy scenarios, such as differing acceptance rates of circular-economy parts. Third, it provides an insurer-oriented decision tool that quantifies trade-offs and co-benefits between emissions reduction and cost containment, supporting the design of claims-management policies consistent with net-zero and circularity strategies. We aim to determine whether and when carbon footprint and economic goals are aligned.

The remainder of the paper is organised as follows: Section~\ref{section avec l'analyse du carbone} introduces the carbon-based risk model and key concepts from operational data. Section~\ref{section d'introduction du modèle avec BINAR(1)} presents the BINAR(1)-based model. Section~\ref{section de propriétés basiques} derives basic statistical properties of both models. Section~\ref{section d'analyse des suites gamm(i,k)} explores the asymptotic behaviour of the claim-count processes used for the proper BINAR(1) model. Section~\ref{section probabilités de ruine} develops the ruin theory results. Finally, Section~\ref{section de simulations de scénarios} provides numerical illustrations and scenario analyses based on real-world data.

\section{Carbon footprint of claim management in motor insurance}
\label{section avec l'analyse du carbone}

We are interested in modeling the carbon footprint of claim management in motor insurance through a bivariate structure. As mentioned in the introduction, we take jointly into account a financial risk and a non-financial risk, which respectively correspond, for a given insurer, to amounts of claims and to quantities of carbon dioxide generated and rejected in order to manage these claims.

In this paper, all the random variables and stochastic processes are defined on a probability space $(\Omega, \mathcal{F}, \p)$. Moreover, $\N = \{0, 1, 2, \dots \}$ is the set of the nonnegative integers, thus \\
$\N^* = \{1,2, \dots \}$ are the positive integers.

\subsection{A bivariate structure for carbon footprint}

To estimate the financial and environmental impacts of automobile collision claims, we rely on two complementary sources: the French \textit{Livre Blanc}, namely the \textit{"White Booklet on carbon footprint in claims management"} (\cite{livreblanc}), and the 2023 Allianz report \textit{"Repair or Replace – Investigating the relative GHG emissions of vehicle part repair vs. replacement"}, \cite{Allianzrapport2023}. 
If the Allianz report rather focuses on the different operations linked with the repairs (including piece replacement or painting), the \textit{Livre Blanc} also evokes the necessary assistance provided to the client who has suffered an accident: for example, by proposing towing, a taxi or a replacement vehicle.

More specifically, we are willing to investigate what can be done and chosen in terms of replacement of a damaged piece, in case the direct repair is difficult or too costly. We can distinguish between two categories: brand-new parts and reused parts from the circular economy (called PIEC in the \textit{Livre Blanc}), which often correspond to pieces still functioning taken from a damaged car dedicated to be destroyed. 
In 2017, the European Union published a list with a quite wide range of car parts for which the use of PIEC is permitted.
Specifically, the use of PIEC is particularly encouraged in France, in particular thanks to the Agec law voted in 2020 (\textit{loi sur l'Antigaspillage pour une économie circulaire}), namely "law on anti-waste for a circular economy". 

Therefore, it entices us to consider a bivariate claim structure, where the $k^{th}$ claim can be represented here as:
\[
Z_k =
\begin{pmatrix}
X_{1,k} \\
X_{2,k}
\end{pmatrix}
+ Y_k
\]
where 
$X_{1,k}$ is seen as the minimum financial cost of the $k^{th}$ claim, and $X_{2,k}$ is the baseline CO$_2$ emissions associated with the claim. Moreover, $Y_k$ is divided into two components, $Y_{1,k}$ and $Y_{2,k}$, which are the respective optional additional cost, or optional additional emissions. The latter two quantities may be equal to 0, depending on the choices made for the repair following the car damage. 

\newpage
\subsection{Different types of correlations}

According to conventional scenarios, a trade-off between financial cost and environmental impact should imply a negative correlation between these two types of risks. Indeed, it seems reasonable to assume that a small fee entails the fact that the insurance company chose a simpler, but less environmentally-friendly solution. As a concrete example, one can think, in terms of travelling, of the difference between using the train 
instead of a plane: it takes more time (and may be more expensive), but the associated carbon emission is expected to be lower.

Thus, in most cases,
\begin{equation}
Y_k \ = \ \left\{
\begin{array}{ll}
\begin{pmatrix}
Y_{1,k} \\
0
\end{pmatrix}, & \text{if carbon footprint reduction is prioritized} \, , \\[1em]
\begin{pmatrix}
0 \\
Y_{2,k}
\end{pmatrix}, & \text{if cost containment is prioritized.}
\end{array}
\right.  
\end{equation}

This representation makes sense when one considers vehicle painting, since we observe a negative correlation between cost and carbon footprint. For instance, using high-quality, water-based paints may result in a higher financial cost but lead to significantly lower emissions compared to cheaper solvent-based paints. In this case, insurers face a true trade-off between economic and ecological outcomes. The \textit{Livre Blanc} specifies that for each collision claim, an average number of 2.31 operations of painting has to be made, representing an average quantity of 39.3 kgCO$_2$ generated per claim, and stands for 26.8 \% of the total emissions for the claims studied by the booklet, so it is a significant part of the recovery process.

This, for example, may also be the case for electric vehicles (more details are provided in Section \ref{sous-section véhicules elec}).
\medskip

However, contrary to the mentioned conventional scenarios, our empirical application to the case of piece replacements and the information we were able to collect reveals the opposite: the additional risk components are positively correlated, that is to say, opting for new parts instead of recycled pieces increases both the financial cost and the carbon emission.

In that case, we shall write instead 
\begin{equation}
Y_k \ = \ \left\{
\begin{array}{ll}
\begin{pmatrix}
Y_{1,k} \\
Y_{2,k}
\end{pmatrix}, & \text{if new pieces are chosen} \, , \\[1em]
0 \, , & \text{if only recycled pieces are chosen.}
\end{array}
\right.  
\end{equation}

Actually, this feature not only simplifies the strategic decision for insurers for this specific operation but also reinforces the dual benefit — economic and ecological — of promoting the adoption of reused parts. Of course, it remains meaningful for any other kind of operation where we could observe such a positive correlation.

Piece replacements represent an average quantity of 37.5 kgCO$_2$ generated per claim (36.2 kgCO$_2$ for the new pieces, and 1.3 kgCO$_2$ for the PIEC), and stand for 25.6 \% of the total emissions for the claims studied by the \textit{Livre Blanc}, so it is also an important part of the problem. 
\medskip

This decomposition allows us further (in Section \ref{section de simulations de scénarios}) to simulate strategic scenarios based on client acceptance rates of PIEC, modelling the trade-off between financial and environmental performance.

Yet, for now, we incorporate this idea of bivariate structure to derive, in the next paragraph, the writing of a discrete-time risk model to try to represent the financial and ecological aspects of the problem we are interested in.

\subsection{The bivariate carbon model}
\label{sous-section de présentation du modèle de carbone}

We now propose a bidimensional risk model in order to represent several aspects of the carbon footprint for claim management in motor insurance. In particular, we emphasise the concept of piece replacement for collision claims. This can be seen as an extension of a more intuitive model with a BINAR(1) process which is defined below in Section \ref{section d'introduction du modèle avec BINAR(1)}. 
\smallskip

Before that, we define the binomial thinning operation:
\begin{defi}[Operation $\circ$] ~\\
\label{def operation thinning}
Let $X$ be an integer-valued random variable, and $\rho \in [0,1]$ a real number. Then, \\
$\rho \circ X \, = \, \sum\limits_{i=1}^{X} b_i(\rho)$, where the $(b_i(\rho))_{i \in \N}$ are i.i.d. random variables with Bernoulli distribution $\cB(\rho)$, independent of $X$. Hence, $\ \rho \circ X | X \ \sim \ \cB(X, \rho)$. 
\end{defi}

\subsubsection{Presentation of the model}

In the following model, we consider $R^C_{1,n}$ corresponds to a financial line, and $R^C_{2,n}$ to a quantity of carbon emitted. We note some variables of the problem with "C" as an exponent to insist on the difference with Equation \ref{équation de définition du modèle en 2D avec BINAR(1)}.

$v_1 \geq 0$ is the initial financial surplus of the insurance company, and $d_1 >0$ is the constant premium rate.

Then, we can see $v_2$ as an initial total quantity of carbon which should not be exceeded by the insurance company over a given period of time, balanced by an additional quantity $d_2$ allowed per period. Thus, we set up or define a kind of carbon liability threshold.
Depending on the context, it could be possible to assume a proportional relation of the type $d_2 = \nu.d_1$, and thus see $\nu >0$ as a quantity of carbon allowed per financial unit of premium, or, approximately, it corresponds to the idea of a quantity of carbon allowed per client insured.

\begin{equation}
\label{modèle avec BINAR(1) pour carbone}
\left(\begin{array}{l}
R^C_{1, n} \\
R^C_{2, n}
\end{array}\right) \ = \ \left(\begin{array}{l}
v_{1} \\
v_{2}
\end{array}\right) 
\, + \, 
n\left(\begin{array}{l}
d_{1} \\
d_{2}
\end{array}\right) 
\, - \, 
\sum\limits_{k=1}^{n} \sum\limits_{i=1}^{N^C_{k}} \left(\begin{array}{c}
 X_{1,k,i} \\
 X_{2,k,i}
\end{array}\right) 
\, - \, 
\sum\limits_{k=1}^{n}\left(\begin{array}{c}
\sum\limits_{i=1}^{M^C_{1,k}} Y_{1,k,i} \\
\sum\limits_{i=1}^{M^C_{2,k}} Y_{2,k,i}
\end{array}\right) \ ,
\end{equation}
where, for $k \geq 1$,
\begin{equation}
\label{equation M^C_k comptage by-claims modèle carbone}
    \left\{
\begin{array}{ll}
        M^C_{1,k} \, = \, \omega_{12} \circ  N^C_{k} \, + \, \delta \circ N^C_{k-1} \, + \, O_k \ ,
     & \\ M^C_{2,k} \, = \, \omega_{21} \circ  N^C_{k} \, +  \, O_k \ ,
     & \\ \text{and,} \ \ O_k = \theta \circ O_{k-1} + \beta \circ N^C_{k-1} \, , \ \ \text{for} \ k \geq 2 \ .
\end{array}
\right.
\end{equation}
At time $k$, all the thinning operations are assumed to be made conditionally independently. Moreover, $\theta, \delta, \beta, \omega_{12}, \omega_{21} \in [0,1]$. In relation to the possible concrete meaning of this model, we display in the next section more precise potential values for some of these parameters.

Actually, the sequence $(O_k)_{k \in \N}$ can be seen as an INAR(1) time serie, where the source term for $k \geq 2$ is $\beta \circ N^C_{k-1}$.

For the $k^{th}$ period, the insurance company suffers $N^C_k$ joint main claims. Comparatively with Equation \ref{équation de définition du modèle en 2D avec BINAR(1)}, here $N^C_k = N_{1,k} = N_{2,k}$. Indeed, an initial claim is always considered to cause both a financial loss and a carbon emission the company is responsible for.
According to what is stated above, the random positive variables $X_{1,k,i}$ and $X_{2,k,i}$ correspond to the baseline claim cost and quantity of CO$_2$ emitted for the $i^{th}$ claim of the $k^{th}$ period. The $(N^C_k)_{k \in \N}$ are independent and identically distributed (i.i.d.) such that $N^C_1 \sim \cP(\lambda), \lambda >0$. 

Then, the model also presents subsequent or additional losses for each line, counted respectively by $M^C_{1,k}$ and $M^C_{2,k}$ at period $k$. The corresponding positive amounts of these by-claims are $(Y_{1,k,i})_{(k,i) \in \N^2}$ and $(Y_{2,k,i})_{(k,i) \in \N^2}$.

As conventions, we take $N^C_{0} \overset{\cL}{=} 0$, $M^C_{1,0} \overset{\cL}{=} 0$, $M^C_{2,0} \overset{\cL}{=} 0$, and 
$O_1 \overset{\cL}{=} 0$.
Moreover, we note $\xi_{1,k} = \E \left(M^C_{1,k} \right)$, and $\xi_{2,k} = \E \left(M^C_{2,k} \right)$ for $k \geq 1$.
We also assume all the loss variables have respective finite means, $\mu_{X_1} >0$, $\mu_{X_2} >0$, $\mu_{Y_1} >0$ and $\mu_{Y_2} >0$.

We assume all the sequences of the different types of losses, $(X_{1,k,i})_{(k,i) \in \N^2}$, $(X_{2,k,i})_{(k,i) \in \N^2}$, $(Y_{1,k,i})_{(k,i) \in \N^2}$ and $(Y_{2,k,i})_{(k,i) \in \N^2}$, are mutually independent and each one i.i.d. Moreover, they are each one independent of the sequence $\left(N^C_k \right)_{k \in \N}$. As a direct consequence, they are also independent of $\left(M^C_k \right)_{k \in \N}$.
\bigskip

To analyse the risks embodied by this model, we are going to apply a classical ruin study, despite the presence of the non-financial risk.

Actually, it makes sense, because it enables to identify risky situations where the insurance company has suffered too high financial losses, or has emitted too much carbon dioxide regarding the given thresholds, or even both.

As suggested previously, in some situations, the priority of an insurance company is to minimise as much as possible its risks and its losses, and then, the carbon footprint is correspondingly reduced if possible. Yet, the following ruin study should spur the insurers to take care of both jointly. We precise later in Section \ref{section d'introduction du modèle avec BINAR(1)} the definitions of these ruin probabilities.

For that, we note $W_n = \sum\limits_{k=1}^n \left(\sum\limits_{i=1}^{N^C_k} (X_{1,k,i} + X_{2,k,i}) +  \sum\limits_{i=1}^{M^C_{1,k}} Y_{1,k,i} + \sum\limits_{i=1}^{M^C_{2,k}} Y_{2,k,i}\right)$, for $n \geq 1$. It corresponds to the sum of the aggregate amount of financial losses until the $n^{th}$ period, and the aggregate quantity of CO$_2$ emitted in relation to all the claims managed during the same time.

\subsubsection{Meaning of the model}

Regarding the model from Equation \ref{équation de définition du modèle en 2D avec BINAR(1)} on a mathematical point of view, the crossing coefficients (here, $\omega_{12}$ and $\omega_{21}$) do not have the same role, even if it looks quite similar. Yet, the main thing here is the fact that $\alpha_{11}$ and $\alpha_{22}$ have been merged through the parameter $\theta$ in a joint thinning operation which involves both lines. It is why, here, it would be awkward to propose a representation with matrices as in Section \ref{section d'introduction du modèle avec BINAR(1)}.

Let us give more concrete details to explain the possible roles of the parameters.

The crossing coefficients $\omega_{12}$ and $\omega_{21}$ allow in particular to take into account for each line the additional costs discussed previously if environmentally-friendly decisions are taken or not.

Regarding $\delta$, we consider that a client who has suffered an accident and has been convinced by his insurer to accept the use of at least one PIEC to repair his vehicle will benefit from a discount on his premium amount during the next period to reward this ecological choice. It could depend on the number or the type of PIEC used, but we can assume this discount is between 10 and 25 \%, and it is not granted if only new pieces are used. 
However, the client may as well withdraw his contract, and in that case, the discount does not have to be made, of course. 
We reckon the proportion of withdrawal at a given period is between 10 and 20\%.

Hence, we may propose the broad estimation $\delta \in [0.3, 0.7]$.
Furthermore, this action has no ecological effects, therefore there is no relapse associated in terms of carbon emission. 

Another possible interpretation of such a coefficient would be a financial penalty due to too high carbon emissions to manage the claims.
\medskip

Besides, the use of PIEC may lead to legal problems for the company: indeed, if there is a problem with such a reused piece after that, the insurer could be assumed liable or responsible for an incorrect repair. An example is the case of a door which no longer opens properly after some time.
Then, this PIEC should be replaced by a more functional piece, which engenders a new couple of financial and ecological costs.
In that case, we could assume a brand-new piece would replace this defective piece, but it is not the most important. 
We can reckon the probability of this type of event is between 5 and 20 \%.
However, if only new pieces are used for the repair, and if such a failure happens, we can assume the client will sue the garage or the repair company instead of the insurance company.

Anyway, whether PIEC or new pieces are used, there must be a one-year (or one-period) guarantee in case something happens to the replaced piece, and another repair has to be made.
This new claim includes not only the necessary assistance but also the procedure of repair.
Then, 
we assume that each time the guarantee is used, it is subsequently replicated for another period. As a detail, if the guarantee is used several times in a period, we consider all these uses to be merged into a single loss for each line.

All these possibilities justify this potential propagation of relapses through by-claims, and are gathered in the coefficients $\beta$ and $\theta$.
We consider $\theta \leq \beta$, since we can imagine that for the subsequent times, the persistent problems will be treated more carefully in order to avoid another couple of by-claims.
We can reckon $\beta \in [0, 0.2]$. 
\bigskip

The model presented above is, of course, not universal, and alternative models may be proposed for the carbon footprint of the claim management process in motor insurance. It is therefore desirable to study not only the properties of this specific model, but also the ones of a more general class of closely related models obtained as a natural extension of existing ones, allowing for either positive or negative correlations between financial and ecological costs. Very often, it is possible to compare the results or adapt the proofs of this general class of models to specific variations of models that take into account carbon footprint specificities. 

Before analysing those bivariate models in parallel, we now briefly present an INAR(1) risk model in the univariate setting, and its natural bivariate extension using a BINAR(1) process, which refers to the general class of models mentioned just above.

\newpage
\section{A bidimensional risk model using a BINAR(1) process}
\label{section d'introduction du modèle avec BINAR(1)}

\subsection{An unidimensional model with an INAR(1) process}

Let us first consider the following unidimensional discrete-time risk model, considered in \cite{minier2026risk}:
\begin{equation}
\label{modèle classique en 1D avec BC et INAR(1) (pour chap 2D)}
    R_n \ = \ u \ + \ c \, n \ - \ \sum\limits_{k=1}^n \sum\limits_{i=1}^{N_k} X_{k,i} \ - \ \sum\limits_{k=1}^n \sum\limits_{i=1}^{M_k} Y_{k,i} \qquad \text{for} \ \, n \in \N \ ,
\end{equation}
where $u \geq 0$ is the initial surplus of the insurance company, and $c >0$ is the constant premium rate.
The i.i.d. sequences of positive random variables $(X_{k,i})_{(k,i) \in \N^2}$ and $(Y_{k,i})_{(k,i) \in \N^2}$ represent respectively the amounts of the main claims and the by-claims. Namely, the company suffers $N_k$ main claims and $M_k$ by-claims at each period $k$, $k \geq 1$, where the sequence $(N_k)_{k \in \N^*}$ is i.i.d., with $N_1 \sim \cP(\lambda)$, $\lambda >0$, and for $k \geq 1$, $M_k$ is expressed in function of $M_{k-1}$, $N_k$ and a constant parameter $\rho \in [0,1[$ according to an INAR process of order 1: 
\begin{equation}
M_k \ = \ \rho \circ M_{k-1} \, + \, N_k \ .
\end{equation}
$\circ$ is the binomial thinning operator, as defined previously.
As an initial condition, $M_0 \sim \cP(\gamma_0)$, with $\gamma_0 >0$.

What is more, in this model, the sequences $X$, $Y$ and $N$ are mutually independent, and $X$ and $Y$ are independent of the sequence $M$, while the relation of dependence between $N$ and $M$ is stated as above.

On a practical viewpoint, this structure with an INAR(1) process implies that during the $k^{th}$ period, each main claim generates directly one by-claim precisely, and then, each by-claim which has occurred at this period engenders with probability $\rho$ a new by-claim for the $(k+1)^{th}$ period.
\medskip

Moreover, as a technical remark, the by-claims have only \textit{positive amounts}: it is thus a simplification of what could happen in practice. Indeed, in the context of cost revising for an insured loss, some main claims may be revised upward or downward. Actually, they are revised downward if some damages appear to be less costly than initially expected, and in that case, we should observe a by-claim with a negative amount.

Nonetheless, considering the precise meaning we assign to our carbon model, it is not a problem in this paper, because the by-claims stand for additional costs rather than cost revision.

\subsection{The classical BINAR(1) risk process}

It is possible to extend the previous univariate discrete-time risk model to deduce the following bivariate risk model of the form, for $n \in \N$:
\begin{equation}
\label{équation de définition du modèle en 2D avec BINAR(1)}
\left(\begin{array}{l}
R_{1, n} \\
R_{2, n}
\end{array}\right) \ = \ \left(\begin{array}{l}
u_{1} \\
u_{2}
\end{array}\right) 
\, + \, 
n\left(\begin{array}{l}
c_{1} \\
c_{2}
\end{array}\right) 
\, - \, 
\sum\limits_{k=1}^{n}\left(\begin{array}{c}
\sum\limits_{i=1}^{N_{1,k}} X_{1,k,i} \\
\sum\limits_{i=1}^{N_{2,k}} X_{2,k,i}
\end{array}\right) 
\, - \, 
\sum\limits_{k=1}^{n}\left(\begin{array}{c}
\sum\limits_{i=1}^{M_{1,k}} Y_{1,k,i} \\
\sum\limits_{i=1}^{M_{2,k}} Y_{2,k,i}
\end{array}\right) \ ,
\end{equation}
where, for $k \geq 1$, the two-dimension vector $M_k = (M_{1,k}, M_{2,k})$ is ruled according to a BINAR(1) process: 
\begin{equation}
    \left\{
\begin{array}{ll}
        M_{1,k} \, = \, \alpha_{11} \circ  M_{1,k-1} \, + \, \alpha_{12} \circ M_{2,k-1} \, + \, N_{1,k} \ ,
     & \\ M_{2,k} \, = \, \alpha_{21} \circ M_{1,k-1} \, + \, \alpha_{22} \circ M_{2,k-1} \, + \, N_{2,k} \quad .
\end{array}
\right.
\end{equation}
Namely, written as matrices,
\begin{equation}
\begin{pmatrix} M_{1,k} \\ M_{2,k} \end{pmatrix} = \begin{pmatrix}
\alpha_{11} & \alpha_{12}  \\
\alpha_{21} & \alpha_{22}
\end{pmatrix} \circ \begin{pmatrix} M_{1,k-1} \\ M_{2,k-1} \end{pmatrix} + \begin{pmatrix} N_{1,k} \\ N_{2,k} \end{pmatrix} \ . 
\end{equation}
\smallskip

\begin{itemize}
    \item $(R_{1,n})_{n \in \N}$ and $(R_{2,n})_{n \in \N}$ stand for two different lines of business of an insurance company.
    \item $u = (u_1, u_2)$ is the vector of initial surpluses, $u_1, u_2 \geq 0$.
    \item $c = (c_1, c_2)$ are the constant premium rates, $c_1, c_2 >0$.
    \item For some integers $1 \leq i,j \leq 2$, $\, \alpha_{ij} \in [0,1[$ is a constant parameter, and $\circ$ is the binomial thinning operator (already defined above).
        \item For a fixed $k \in \N$, and for $1 \leq i,j \leq 2$, the variables $\alpha_{ij} \circ M_{j,k}$ are conditionally independent.
    \item For the $j^{th}$ line ($j \in \{1,2\}$), the company suffers $N_{j,k}$ main claims and $M_{j,k}$ by-claims at each period $k \geq 1$, where the sequence $(N_{j,k})_{k \in \N^*}$ is i.i.d, with $N_{j,1} \sim \cP(\lambda_j)$, $\lambda_j >0$.
    \item We settle $M_{1,0} \sim \cP(\gamma_{1,0})$, and $M_{2,0} \sim \cP(\gamma_{2,0})$, where $\gamma_{1,0} >0$, and $\gamma_{2,0} >0$ are fixed. 
    Moreover, for a fixed $k \geq 1$, we assume that the 2D vector $N_k$ is independent of $M_0, \dots, M_{k-1}$, and we note $\gamma_{j,k} = \E(M_{j,k})$ for $k \in \N$ and $1 \leq j \leq 2$. We prove later that with such assumptions, actually $M_{j,k} \sim \cP(\gamma_{j,k})$.
    \item For the $j^{th}$ line, $(X_{j,k,i})_{(k,i) \in \N^2}$ and $(Y_{j,k,i})_{(k,i) \in \N^2}$ are i.i.d. sequences representing respectively the main claims and by-claims, with respective finite mean $\mu_{X_j}$ and $\mu_{Y_j}$. We may note by $X_j$ and $Y_j$ two variables such that $X_{j,1,1} \sim X_j$, and $Y_{j,1,1} \sim Y_j$.
    \item The sequences $(X_{1,k,i})_{(k,i) \in \N^2}$, $(X_{2,k,i})_{(k,i) \in \N^2}$, $(Y_{1,k,i})_{(k,i) \in \N^2}$, $(Y_{2,k,i})_{(k,i) \in \N^2}$,  $(N_{1,k})_{k \in \N^*}$ and $(N_{2,k})_{k \in \N^*}$ are mutually independent. Moreover, the sequences of vectors $(X_k)_{k \in \N}$ and $(Y_k)_{k \in \N}$ are independent of the sequence $(M_k)_{k \in \N}$. 
    Eventually, the relation of dependence between the sequences of 2D vectors $N$ and $M$ is stated as above through the BINAR(1) formulation.
\end{itemize} 

Here again, we assume which could be seen as a simplification of the reality, that is to say, the by-claims have only \textit{positive amounts}.
\bigskip

From a mathematical viewpoint, this structure with a BINAR(1) process engenders dependence between the two lines of business while representing the possibility of having subsequent losses over several periods coming from the same initial event. More specifically, it implies that during a given period, each main claim suffered in a given line generates directly a by-claim in the same line. Then, other losses may occur from this in the following periods, and it could affect either of the lines. The corresponding probabilities that such further claims happen 
are represented by the coefficients $\alpha_{ij}$.
Moreover, a BINAR(1) process corresponds to a discrete sequence of random variables, and not a continuous process; therefore, it seemed more natural to us to implement such a process in discrete time, all the more since it is one of the first uses of such a time series in ruin theory. Nonetheless, the transcription of such a crossing dependence in continuous time looks possible.
Besides, to our knowledge, there is from now no continuous-time risk model in the ruin theory literature which involves an integer-valued time series as a process counting the claims.

What is more, from a concrete point of view, without delving too much into the details, such a model could be meaningful in a general case of motor insurance: indeed, the lines could stand for the material part, the civil liability part, or the health damage part, for instance. In such a concrete context, we would observe dependence between the lines, in particular an event in a given line could engender repercussions or subsequent claims on both lines.
\bigskip

The model is said to be independent if $\alpha_{12} = \alpha_{21} = 0$. However, if we keep the independence assumptions between the claims, the two lines of business will act separately, as if they were ruled by two respective independent INAR(1) processes (one could refer to \cite{minier2026risk} for corresponding results). Therefore, such a study would be interesting only if we modified the previous assumptions.
\bigskip

At time $n \geq 1$, we can define the aggregate amount of the losses: \\ 
$S_n = \sum\limits_{k=1}^n \left( \sum\limits_{i=1}^{N_{1,k}} X_{1,k,i} +
\sum\limits_{i=1}^{N_{2,k}} X_{2,k,i} + \sum\limits_{i=1}^{M_{1,k}} Y_{1,k,i} +
\sum\limits_{i=1}^{M_{2,k}} Y_{2,k,i} \right)$.
\bigskip

Eventually, there are several ways to characterise the ruin of a bidimensional risk process.
First, we set the stopping times:
$\tau_1(u_1) = \inf \{ n \in \N^* \ / \ R_{1,n} < 0 \, | \, R_{1,0} = u_1 \}$, and \\
$\tau_2(u_2) = \inf \{ n \in \N^* \ / \ R_{2,n} < 0 \, | \, R_{2,0} = u_2 \}$.
Thus, $\tau_1(u_1) , \, \tau_2(u_2) \ \in \ \N^* \cup \{+ \infty\}$.

Then, we define the ruin probabilities:
\begin{itemize}
    \item $\psi_{1}(u_1) = \p(\tau_{1}(u_1) < \infty)$, and the same thing for $\psi_2(u_2)$.
    \item $\psi_{and}(u) = \p(\tau_{and}(u) < \infty)$, where $\tau_{and}(u) = \max\{ \tau_1(u_1), \tau_2(u_2) \}$.
    \item $\psi_{or}(u) = \p(\tau_{or}(u) < \infty)$, where $\tau_{or}(u) = \min \{ \tau_1(u_1), \tau_2(u_2) \}$.
    \item $\psi_{sum}(u) = \p(\tau_{sum}(u) < \infty)$,   where $\tau_{sum}(u) = \inf \{ n \in \N^* \ / \ R_{1,n} + R_{2,n} < 0 \}$.
    \item For an integer $T \geq 1$, we can define $\pSn = \p(\tau_{sum}(u) \leq T)$, and the same thing for $\pOn(u)$, $\pAn(u)$, $\psi^T_{1}(u_1)$, or $\psi^T_{2}(u_2)$.
\end{itemize}

As direct consequences, we get $\, 0 \, \leq \, \psi_{and}(u) \, \leq \, \psi_1(u_1) , \, \psi_2(u_2)$, \\ and then,
$0 \, \leq \,\psi_1(u_1) , \, \psi_2(u_2)  , \, \psi_{sum}(u) \, \leq \, \psi_{or}(u) \, \leq \, \psi_1(u_1) + \psi_2(u_2) \, $.

What is more, we would have the same inequalities by looking at a finite-time horizon.

It is also possible to define the same kind of ruin probabilities for the model from Equation \ref{modèle avec BINAR(1) pour carbone}, such as $\pSC(v)$, $\psi_1^C(v_1)$, or $\psi_2^C(v_2)$.

\section{Basic properties}
\label{section de propriétés basiques}

In this section, we outline some basic properties related to the two types of models. 
The main mathematical results here are the covariances or the net profit conditions. As we will explain it, the subsequent and underlying aim of such results is to determine if whether or not, we are already able to state if the ruin is almost sure.

What is more, as specified in Section \ref{sous-section de présentation du modèle de carbone}, we provide a study involving both financial and ecological risks jointly, but we give more details about this aspect at the beginning of Section \ref{section probabilités de ruine}, which corresponds to the proper ruin probabilities results.

\subsection{Counting processes of the by-claims}

Before providing the net profit conditions related to both models, we yield some properties about the distributions of the counting processes linked with the by-claims for each model.
\medskip

First, we outline a useful lemma:

\begin{lem}
\label{lemme gamma_n}
Let $(a_n)_{n \in \N}$ be a sequence satisfying $a_0 >0$, and $a_{n+1} = \rho .a_n + b$ for $n \in \N$, where $\rho \in [0,1[$ and $b >0$.
Then, $a_n = \rho^n \left(a_0 - \frac{b}{1 - \rho} \right) + \frac{b}{1 - \rho}$, the sequence is monotonic, and it converges towards $\frac{b}{1 - \rho}$.
\end{lem}

\begin{proof}
The expression may be easily checked by induction, then the monotony and the limit follow from this expression, and the fact that $|\rho| < 1$.
\end{proof}
\smallskip

\begin{prop}[Values of the limits] ~\\
\label{Valeur de la limite des suites gamma(i,k)}
When the sequences $(\gamma_{i,k})_{k \in \N, i \in \{1,2\}}$ converge, their limits are 
\begin{equation}
\left \{
\begin{array}{ll}
        \gamma_{1,k} \ \overset{k \to \infty}{\longrightarrow} \ \ \gamma_{1,\infty} \, = \, \frac{(1- \alpha_{22})  \lambda_1 + \alpha_{12} \lambda_2}{(1 - \alpha_{11})(1 - \alpha_{22}) - \alpha_{12} \alpha_{21}} \quad ,
     & \vspace{0.3 cm} \\  \gamma_{2,k} \ \overset{k \to \infty}{\longrightarrow} \ \ \gamma_{2,\infty} \, = \, \frac{(1- \alpha_{11})  \lambda_2 + \alpha_{21} \lambda_1}{(1 - \alpha_{11})(1 - \alpha_{22}) - \alpha_{12} \alpha_{21}} \quad .
\end{array}
\right.
\end{equation}
Specifically, these limits do not depend on the initial values $\gamma_{1,0}$ and $\gamma_{2,0}$.
\end{prop}

\begin{proof}
As the thinning operator $\circ$ satisfies $\E(\alpha \circ Z) = \alpha \E(Z)$ for $\alpha \in [0,1]$, and an integer-valued random variable $Z$, we get, for $k \in \N$,
\begin{equation}
\label{relation de recc gamma(i,n) en 2D}
\left\{
\begin{array}{ll}
        \gamma_{1,k+1} = \lambda_1 + \alpha_{11} \, \gamma_{1,k} + \alpha_{12} \, \gamma_{2,k} \ ,
     & \\ \gamma_{2,k+1} = \lambda_2 + \alpha_{21} \, \gamma_{1,k} + \alpha_{22} \, \gamma_{2,k} \ .
\end{array}
\right.
\end{equation}
By writing these relations satisfied for $k \to \infty$, it yields the expressions outlined.
\end{proof}

\medskip

For the model from Equation \ref{modèle avec BINAR(1) pour carbone}, from the fact that $\left(N^C_k\right)_{k \in \N}$ is i.i.d., we deduce directly that $O_k \in \sigma(N_0, \dots, N_{k-1})$, so $O_k$ is independent of $N^C_j$, for $j \geq k$ (in particular from $N^C_k$). This is helpful to deduce the following proposition:

\begin{prop}[Distribution of the $(O_k)_{k \in \N}$]
\label{prop distribution des O_k}
For $k \geq 2$, we have $O_k \sim \cP(\zeta_k)$, where
$$\zeta_k \, = \, \frac{\beta \lambda}{1 - \theta} (1 - \theta^{k-1}) \ .$$
In particular, $O_k \overset{\cL}{\to} O_\infty$, where $O_\infty \sim \cP \left( \frac{\beta \lambda}{1 - \theta}\right)$.
\end{prop}

To prove that, one may simply adapt properly Lemma \ref{lemme gamma_n} and use the fact that the $(N^C_k)_{k \in \N}$ are i.i.d.
\medskip

However, because of the underlying dependence in the expression of $M^C_{1,k}$ regarding the involvement of $N^C_{k-1}$, we cannot deduce that the variables $(M^C_{1,k})_{k \in \N}$ are also Poisson distributed. 

However, it is possible to deduce the $(M^C_{2,k})_{k \in \N}$ are Poisson distributed, similarly as for the $(O_k)_{k \in \N}$

In all cases, it is possible to explicitly state their means:

\begin{cor}[Distribution of the $(M^C_{1,k})_{k \in \N}$ and $(M^C_{2,k})_{k \in \N}$]
By traducing Equation \ref{equation M^C_k comptage by-claims modèle carbone} and using Proposition \ref{prop distribution des O_k}, we get, for $k \geq1$:
\begin{equation}
\xi_{1,k} \, = \, \lambda \left( \frac{(\omega_{12} + \delta)(1 - \theta) + \beta(1 - \theta^{k-1})}{1 - \theta} \right) \, , 
\end{equation}
and
\begin{equation}
\xi_{2,k} \, = \, \lambda \left( \frac{\omega_{21}(1 - \theta) + \beta(1 - \theta^{k-1})}{1 - \theta} \right) \ .
\end{equation}
In particular, they are both nondecreasing, and $M^C_{2,k} \sim \cP(\xi_{2,k})$. \\
Moreover, when $k \to \infty$, $\xi_{1,k} \to \xi_{1,\infty}$, and $\xi_{2,k} \to \xi_{2,\infty}$, where
\begin{equation}
\xi_{1,\infty} \, = \, \lambda \left( \frac{(\omega_{12} + \delta)(1 - \theta) + \beta}{1 - \theta} \right) \, , \ \text{and} \ \
\xi_{2,\infty} \, = \, \lambda \left( \frac{\omega_{21}(1 - \theta) + \beta}{1 - \theta} \right) \ .
\end{equation}
\end{cor}

\begin{rem}[Comparison of the respective counting processes of the by-claims] ~\\
We can see these last expressions are quite different from the expressions found in Proposition \ref{Valeur de la limite des suites gamma(i,k)}. It is because in the autoregressive expressions, there are joint components coming from $\left(N^C_k \right)_{k \in \N}$ and $(O_k)_{k \in \N}$, but no component coming directly from $(M^C_{1,k})_{k \in \N}$ or $(M^C_{1,k})_{k \in \N}$, contrary to the BINAR(1) process.  

Furthermore, for the bivariate carbon risk model, we are sure the sequences $(\xi_{1,k})_{k \in \N}$ and $(\xi_{2,k})_{k \in \N}$ converge, whereas for the classical model with the BINAR(1) process, it is not sure $(\gamma_{1,k})_{k \in \N}$ and $(\gamma_{2,k})_{k \in \N}$ converge. Numerous conditions regarding the possible convergence of these last two sequences are given in Section \ref{section d'analyse des suites gamm(i,k)}.
\end{rem}

\subsection{Net profit conditions}

After these preliminaries, we are able to outline more precisely the net profit conditions related to each model.
\smallskip

These conditions are equivalent to having $\E(R_{1,n} - u_1) >0$ and $\E(R_{2,n} - u_2) >0$ for every $n \in \N^*$ for the classical BINAR(1) process (and similarly for the bivariate carbon model).

\begin{prop}[Net profit condition for the BINAR(1) model]~\\
\label{CSprofitneten2D}
Let us assume that the sequences $(\gamma_{1,k})_{k \in \N}$ et $(\gamma_{2,k})_{k \in \N}$ converge. Thus, the net profit condition can be expressed as: 
\begin{equation}
\label{eq profit net pour le modèle BINAR(1)}
    \left\{
\begin{array}{ll}
    c_1 > \lambda_1 \mu_{X_1} + \mu_{Y_1} . \underset{k \in \N^*}{\sup} (\gamma_{1,k}) \ , \quad \text{and}
     & \\ c_2 > \lambda_2 \mu_{X_2} + \mu_{Y_2} . \underset{k \in \N^*}{\sup} (\gamma_{2,k}) \ .
\end{array}
\right. 
\end{equation}

If the sequences do not converge, then the net profit condition is not fulfilled. We refer to Section \ref{section d'analyse des suites gamm(i,k)} for more details about such conditions of convergence. \\
Actually, this is a general formulation which distinguishes between the two lines of business. Yet, if we focus on the ruin probability $\pS(u)$, we should rather write it: \\
$c_1 + c_2 \, > \, \lambda_1 \mu_{X_1} + \lambda_2 \mu_{X_2} + \mu_{Y_1} . \underset{k \in \N^*}{\sup} (\gamma_{1,k}) + \mu_{Y_2} . \underset{k \in \N^*}{\sup} (\gamma_{2,k})$.
\end{prop}
\smallskip

In particular, if the last condition is not satisfied, then the ruin is almost sure, $\pS(u) = 1$. If in Equation \ref{eq profit net pour le modèle BINAR(1)}, the condition is not satisfied for a given line $i \in \{1,2\}$, then $\psi_i(u_i) = 1$.

Therefore, in the remainder of the paper, these conditions are always assumed to be true.
\bigskip

\newpage
Then, regarding the bivariate carbon model:

\begin{prop}[Net profit condition for the carbon model]~\\
\label{profit net du modèle de carbone}
The net profit condition for this model is
\begin{equation}
    \left\{
\begin{array}{ll}
    d_1 > \lambda \left( \mu_{X_1} + \mu_{Y_1}  \frac{(\omega_{12} + \delta)(1 - \theta) + \beta}{1 - \theta} \right)  \ , \quad \text{and}
     & \\ d_2 > \lambda \left( \mu_{X_2} + \mu_{Y_2}  \frac{\omega_{21}(1 - \theta) + \beta}{1 - \theta} \right)  \quad .
\end{array}
\right. 
\end{equation}
Specifically, regarding the ruin probability $\pS(u)$, we could rather write it:
$$d_1 + d_2 \, > \, \lambda \left( \mu_{X_1} + \mu_{X_2} \, + \, \mu_{Y_1} \frac{(\omega_{12} + \delta)(1 - \theta) + \beta}{1 - \theta} \, + \, \mu_{Y_2}  \frac{\omega_{21}(1 - \theta) + \beta}{1 - \theta} \right) \ . $$
\end{prop}

\begin{proof}
As before, the expression is $d_i > \lambda \mu_{X_i} + \mu_{Y_i} .\underset{k \geq 1}{\sup} \left(\xi_{i,k} \right)$ for each line \\
$i \in \{1,2\}$. Since $(\xi_{i,k})_{k \geq 1}$ is nondecreasing, we get $\underset{k \geq 1}{\sup} \left(\xi_{i,k} \right)  = \xi_{i,\infty}$.
\end{proof}

If these conditions are not fulfilled, then the corresponding ultimate ruin probabilities are once again equal to 1.
\bigskip

Then, it is possible to compare the net profit conditions of the two types of models.

If we consider the values of the means $\mu_{X_1}$, $\mu_{Y_1}$, $\mu_{X_2}$ and $\mu_{Y_2}$ are fixed, it is interesting to compare these conditions to see if one case is stricter than the other.

For that, we would like to compute the differences $\gamma_{i,\infty} - \xi_{i,\infty}$, and we have to adapt the formalism: here, we consider that $\lambda = \lambda_1 = \lambda_2$, $\alpha_{12} = \omega_{12}$, $\alpha_{21} = \omega_{21}$, and $\theta = \alpha_{11} = \alpha_{22}$. Moreover, to simplify, we choose to take $\beta = \theta$.

The computation of $\gamma_{2,\infty} - \xi_{2,\infty}$ yields:
$$
\gamma_{2,\infty} - \xi_{2,\infty} = \lambda \left( \frac{(1 - \theta)^3 + \theta \alpha_{12} \alpha_{21} + \alpha_{21}(1 - \theta)(\alpha_{12} \alpha_{21} + 2 \theta - \theta^2)}{(1 - \theta) \left[ (1 - \theta)^2 - \alpha_{12} \alpha_{21} \right]} \right) \ ,
$$
which shows that $\gamma_{2,\infty} > \xi_{2,\infty}$, as $\theta \leq 1$.

The difference with the first line is the involvement of the parameter $\delta$:
$$
\gamma_{1,\infty} \, - \,  \xi_{1,\infty} \, =  \, \lambda \, \frac{(1 - \theta)^3 (1 - \delta) + \theta \alpha_{12} \alpha_{21} +  + \alpha_{12}(1 - \theta)(\alpha_{12} \alpha_{21} + 2 \theta - \theta^2) + \delta(1 - \theta)\alpha_{12} \alpha_{21}}{(1 - \theta) \left[ (1 - \theta)^2 - \alpha_{12} \alpha_{21} \right]} \, ,  
$$
Since $\delta \leq 1$, we get $\gamma_{1,\infty} \, - \, \xi_{1,\infty} >0$.
\bigskip

To summarise it as a result,
\begin{prop}
\label{prop comparaison conditions de profit net}
If $c_1$ and $c_2$ satisfy the net profit condition for the BINAR(1) model, if $d_1 \geq c_1$ and $d_2 \geq c_2$, the net profit condition is also fulfilled for the bivariate carbon model.

It still stands if we assume more generally that $\lambda \leq \lambda_1, \lambda_2$, $\omega_{12} \leq \alpha_{12}$, $\omega_{21} \leq \alpha_{21}$,
or \\
$\theta \leq \beta \leq \alpha_{11}, \alpha_{22}$, since it increases the risk incurred by the classical model compared to the carbon model.
\end{prop}

\subsection{Covariances and correlations}

For the classical BINAR(1) model, as the counting processes of the main claims are Poisson distributed, we can deduce the variances: for $k \geq 1$, we have $\V(N_{1,k}) = \lambda_1$, and $\V(N_{2,k}) = \lambda_2$.

Then, for the general covariances, it is difficult to get them explicitly as soon as a variable $M_{j,k}$ is involved, for $1 \leq j \leq 2$, and $k \geq 1$, so instead we present recurrence relations.

Of course, $\cov(N_{i,k}, N_{j,l}) = 0$ as soon as $i \neq j$ or $k \neq l$, but also, \\
$\cov(N_{i,k}, M_{j,0}) = 0 \,$ for every $(i,j) \in \{1,2\}^2$ and $k \geq 1$.

\begin{prop}[Covariances of the counting processes] ~\\
\label{Covariance en 2D entre N et M}
Let us take two integers $n,k \geq 1$.
\smallskip

i) If $k=n$,
$\left\{
\begin{array}{ll}
        \cov(N_{1,k}, M_{1,n}) = \lambda_1 \ ,
     & \\  \cov(N_{1,k}, M_{2,n}) = 0 \ .
\end{array}
\right. $
Otherwise, 
$$
\left\{
\begin{array}{ll}
        \cov(N_{1,k}, M_{1,n+1}) = \alpha_{11} \cov(N_{1,k}, M_{1,n}) + \alpha_{12} \cov(N_{1,k}, M_{2,n}) \ ,
     & \\  \cov(N_{1,k}, M_{2,n+1}) = \alpha_{21} \cov(N_{1,k}, M_{1,n}) + \alpha_{22} \cov(N_{1,k}, M_{2,n}) \ .
\end{array}
\right.
$$

ii) If $k=n$,
$
\left\{
\begin{array}{ll}
        \cov(N_{2,k}, M_{1,n}) = 0 \ ,
     & \\  \cov(N_{2,k}, M_{2,n}) = \lambda_2 \ .
\end{array}
\right.
$
Otherwise, 
$$
\left\{
\begin{array}{ll}
        \cov(N_{2,k}, M_{1,n+1}) = \alpha_{11} \cov(N_{2,k}, M_{1,n}) + \alpha_{12} \cov(N_{2,k}, M_{2,n}) \ ,
     & \\  \cov(N_{2,k}, M_{2,n+1}) = \alpha_{21} \cov(N_{2,k}, M_{1,n}) + \alpha_{22} \cov(N_{2,k}, M_{2,n}) \ .
\end{array}
\right.
$$
\end{prop}

\begin{proof}
The key property for these relations is: for two integer-valued variables $Z_1$ and $Z_2$, and $\alpha \in [0,1]$, $\cov(\alpha \circ Z_1,Z_2) = \alpha \cov(Z_1,Z_2)$.
\end{proof}

\begin{prop}[Covariances of the $(M_{j,k})_{k \in \N, j \in \{1,2\}}$] ~\\
\label{Covariance entre les M(j,k)}
At range $0$, 
$\cov(M_{1,0}, M_{2,0}) = 0$. \\
Then, for $k,n \in \N$:
$$
\left\{
\begin{array}{ll}
        \cov(M_{1,k}, M_{1,n+1}) = \alpha_{11} \cov(M_{1,k}, M_{1,n}) + \alpha_{12} \cov(M_{1,k}, M_{2,n}) \ ,
     & \\  \cov(M_{1,k}, M_{2,n+1}) = \alpha_{21} \cov(M_{1,k}, M_{1,n}) + \alpha_{22} \cov(M_{1,k}, M_{2,n}) \ ,
     & \\ \cov(M_{2,k}, M_{1,n+1}) = \alpha_{11} \cov(M_{2,k}, M_{1,n}) + \alpha_{12} \cov(M_{2,k}, M_{2,n}) \ ,
     & \\  \cov(M_{2,k}, M_{2,n+1}) = \alpha_{21} \cov(M_{2,k}, M_{1,n}) + \alpha_{22} \cov(M_{2,k}, M_{2,n}) \ .
\end{array}
\right.
$$
Even if it is difficult to compute, the variables $M_{1,k}$ and $M_{2,n}$ are clearly positively correlated.
\end{prop}
\smallskip

For the bivariate carbon model, once again, we can infer some results from the fact that some variables are Poisson distributed.

\begin{prop}[Variances for the carbon model] ~\\
\label{Variances en 2D pour le carbone}
For $k \geq 1$, we have $\V(N^C_{k}) = \lambda$, and for $k \geq 2$, $\V(O_k) = \frac{\beta \lambda (1- \theta^{k-1})}{1 - \theta}$. \\
Then, $\V(M^C_{2,k}) = \xi_{2,k} = \lambda \left( \frac{\omega_{21}(1 - \theta) + \beta(1 - \theta^{k-1})}{1 - \theta} \right)$. \\
For $M^C_{1,k}$, $\V(M^C_{1,1}) = \omega_{12} \lambda$, and for $k \geq 2$, $\V(M^C_{1,k}) = (\omega_{12} + \delta) \lambda + \frac{\beta \lambda (1- \theta^{k-1})}{1 - \theta} + \beta \delta \lambda$. \\
\end{prop}

\begin{proof}
For $\V(M^C_{1,k})$, we use the fact that for an integer-valued variable $X$, \\ $\alpha, \beta \in [0,1]$, and if the thinning operations are made conditionally independently, then \\
$\cov(\alpha \circ X, \beta \circ X) = \alpha \beta \, \V(X)$. 
\end{proof}

\begin{prop}[Covariances for the carbon model]~\\
\label{Variances en 2D pour le carbone}
For $k \geq 1$, we have $\cov(M^C_{1,k}, N^C_{k}) = \omega_{12} \lambda$, and $\cov(M^C_{2,k}, N^C_{k}) = \omega_{21} \lambda$. \\
Then, if $k \geq 2$, we get $\cov(M^C_{1,k}, N^C_{k-1}) = (\delta + \beta) \lambda$, and $\cov(M^C_{2,k}, N^C_{k-1}) = \beta \lambda$. \\
If $k \geq 3$, for $2 \leq j \leq k-1$, $\cov(M^C_{1,k}, N^C_{k-j}) = \cov(M^C_{2,k}, N^C_{k-j}) = \beta \theta^{j-1} \lambda$. \\
Eventually, if $j > k$, then $\cov(M^C_{1,k}, N^C_{j}) = \cov(M^C_{2,k}, N^C_{j}) = 0$.
\end{prop}

\begin{prop}[Covariances of the $(M^C_{j,k})_{k \in \N, j \in \{1,2\}}$]~\\
First, $\cov(M^C_{1,1}, M^C_{2,1}) = \omega_{12} \omega_{21} \lambda$. \\
For $k \geq 2$, $\cov(M^C_{1,1}, M^C_{2,k}) = \omega_{12} \beta \theta^{k-2} \lambda$, and $\cov(M^C_{1,k}, M^C_{2,1}) = \omega_{21} \beta \theta^{k-2} \lambda$. \\
Then, for $k \geq 2$, $\cov(M^C_{1,k}, M^C_{2,k}) = (\omega_{12} \omega_{21} + \beta \delta) \lambda + \frac{\beta \lambda (1- \theta^{k-1})}{1 - \theta}$. \\
Now, for $k,j \geq 2$: if $j= k-1$, $\cov(M^C_{1,k}, M^C_{2,j}) \, = \, \omega_{21} (\beta + \delta) \lambda + \frac{\beta \lambda (1- \theta^{k-2})}{1 - \theta} \,$, \\
if $j \leq k-2$, $\cov(M^C_{1,k}, M^C_{2,j}) \, = \, \omega_{21} \beta \theta^{k-j-1} \lambda + \frac{\beta \lambda (1- \theta^{k-j-1})}{1 - \theta} \,$, \\
eventually, if $j > k$, $\cov(M^C_{1,k}, M^C_{2,j}) \, = \,  (\omega_{12} + \theta \delta) \beta \theta^{j-k-1} \lambda  + \frac{\beta \lambda (1- \theta^{j-k-1})}{1 - \theta} \,$.
\end{prop}

\begin{proof}
For $k \geq 2$, by developing, \\
$\cov(M^C_{1,k}, M^C_{2,k}) = \V(O_k) + \cov(\omega_{12} \circ N^C_k, \omega_{21} \circ N^C_k) + \cov (\beta \circ N^C_{k-1}, \delta \circ N^C_{k-1})$.
The other computations are similar.
\end{proof}

Therefore, contrary to the classical BINAR(1) model, for the bivariate carbon model, it is possible to compute explicitly all the covariances of the counting processes. It is mainly due to the fact there is no real crossing dependence between the two lines for this model, we simply have a joint dependence thanks to the variable $O_k$ which is involved in both expressions of $M^C_{1,k}$ and $M^C_{2,k}$.
\bigskip

As a summary, through these covariance results, we discovered and enhanced some differences in behaviour between the two models of this paper.
\bigskip

\section{Asymptotic behaviour of the BINAR(1) process $(M_k)_{k \in \N}$}
\label{section d'analyse des suites gamm(i,k)}

In this section, we study for $i \in \{1,2\}$ the sequences $(\gamma_{i,k})_{k \in \N}$, which as a recall, correspond to the means of the variables $(M_{i,k})_{k \in \N}$.
We display here several properties to characterise their joint behaviour. As a recall, in Proposition \ref{Valeur de la limite des suites gamma(i,k)}, the values of the limits are already specified in case the two sequences converge.

This appears necessary before studying proper ruin probabilities, because it allows to characterise the asymptotic behaviour of the risk process $(R_n)_{n \in \N}$, and identify a potential divergence. In particular, links can be made with the net profit condition (Proposition \ref{CSprofitneten2D}): as precised in the previous section, if there is a divergence, then this condition is not fulfilled (and the ultimate ruin is almost sure). However, it concerns the classical model, but we cannot deduce that it necessarily implies such a risky situation for the carbon model.

\begin{prop}[Equivalent characterizations] ~\\
\label{caracterisations equivalentes suites gamma(i,k)}
i) $(\gamma_{1,k})_{k \in \N}$ is bounded if and only if $(\gamma_{2,k})_{k \in \N}$ is bounded. \\ 
ii) $(\gamma_{1,k})_{k \in \N}$ converges if and only if $(\gamma_{2,k})_{k \in \N}$ converges. \\ 
iii) The two sequences $(\gamma_{1,k})_{k \in \N}$ and $(\gamma_{2,k})_{k \in \N}$ are bounded if and only if they converge.
\end{prop}

\begin{proof}
i) It holds that $\gamma_{2,k+1} \leq \alpha_{21}M_1 + \lambda_2 + \alpha_{22}  \gamma_{2,k} \ $ if $M_1 > 0$ is an upper bound of $(\gamma_{1,k})_{k \in \N}$.
By defining $t_0 = \gamma_{2,0}$, and $t_{k+1} = \alpha_{22}  t_k + \alpha_{21} M_1 + \lambda_2$, we get $\gamma_{2,k} \leq t_k$ for $k \in \N$. Moreover, $(t_k)_{k \in \N}$ converges according to Lemma \ref{lemme gamma_n}, so it is bounded. We deduce that $(\gamma_{2,k})_{k \in \N}$ is bounded, and the proof is similar for the converse.
\medskip

ii) We assume $(\gamma_{1,k})_{k \in \N}$ converges. Let us note $l_1$ its limit. Let $0 < \e \leq 1$ be a real number, and $N_0 \in \N$ such that for $k \geq N_0$, $|\gamma_{1,k} - l_1| < \e$. Then, for $k \geq N_0$,
$$
\alpha_{21} (l_1 - \e) + \alpha_{22}  \gamma_{2,k} + \lambda_2 \leq \gamma_{2,k+1} \leq \alpha_{21} (l_1 + \e) + \alpha_{22} \gamma_{2,k} + \lambda_2 \ .
$$
Let us define $(u_k)_{k \geq N_0}$ and $(v_k)_{k \geq N_0}$ such that:
$\ \left\{
\begin{array}{ll}
        u_{N_0} = v_{N_0} = \gamma_{2,N_0} \ ,
     &  \\ u_{k+1} = \alpha_{21} (l_1 - \e) + \alpha_{22}  u_k + \lambda_2 \ ,
     & \\ v_{k+1} = \alpha_{21} (l_1 + \e) + \alpha_{22}  v_k + \lambda_2 \ .
\end{array}
\right.$
\smallskip

Then, $u_k \leq \gamma_{2,k} \leq v_k$ for $k \geq N_0$. Moreover, the sequences $(u_k)_k$ and $(v_k)_k$ converge thanks to Lemma \ref{lemme gamma_n}, respectively to $\frac{\alpha_{21} (l_1 - \e) + \lambda_2}{1 - \alpha_{22}}$, and $\frac{\alpha_{21} (l_1 + \e) + \lambda_2}{1 - \alpha_{22}}$.

So there exists $N_1 \in \N$ such that for $k \geq N_1 \geq N_0$, we have $\left|u_k - \lim\limits_{k \to \infty} u_k \right| < \e$, and \\
$\left|v_k - \lim\limits_{k \to \infty} v_k \right| < \e$. Then, for $k \geq N_1$,
$$
\frac{\lambda_2 + \alpha_{21}l_1}{1 - \alpha_{22}} - \e \, \frac{2 l_1 \alpha_{21} + \lambda_2}{1 - \alpha_{22}} \ \leq \ \gamma_{2,k} \ \leq \ \frac{\lambda_2 + \alpha_{21}l_1}{1 - \alpha_{22}} + \e \, \frac{\lambda_2 \alpha_{21}(2 l_1 + 1)}{1 - \alpha_{22}} \ .
$$
We deduce that $(\gamma_{2,k})_{k \in \N}$ converge, and its limit is $\frac{\lambda_2 + \alpha_{21}l_1}{1 - \alpha_{22}}$.

The proof would be the same for the converse.
\bigskip

iii) One sense is obvious since a converging sequence is always bounded.

Now, we assume the sequences are bounded.

We consider the matrix $A = \begin{pmatrix}
\alpha_{11} & \alpha_{12}  \\
\alpha_{21} & \alpha_{22}
\end{pmatrix}$ and the vector 
$L = \begin{pmatrix} \lambda_1  \\ \lambda_2
\end{pmatrix}$. Moreover, we note \\
$Z_k = \begin{pmatrix} \gamma_{1,k}  \\ \gamma_{2,k}
\end{pmatrix}$. Then, for $k \in \N$,
$\ Z_{k+1} \ = \ A \, Z_k + L \ = \ A^{k+1} \, Z_0 + \left( \sum\limits_{i=0}^{k} A^i\right)  L \, .$

The aim is to know if the sequence of 2D-vectors $(Z_k)_{k \in \N}$ can be bounded without converging.
\smallskip

Some linear algebra theorems ensure that for a given matrix
$M \in \cM_d(\C) , \, d \in \N^*$, the sequence
$(M^k)_{k \in \N}$ converges if its spectral radius $\displaystyle \rho(M) = \sup \{|\xi|, \, \xi \, \in  \operatorname{Sp}(M)\}$
satisfies \\
$\rho(M) < 1$, and the limit is the matrix 0 in that case. 

Moreover, if $(M^k)_{k \in \N}$ is bounded in $\cM_d(\C)$, it implies that $\rho(M) \leq 1$. If, in addition, $M$ is diagonalizable, then the converse assertion is true.
\smallskip

Therefore, let us focus on the spectrum of $A$. Its characteristic polynomial is:
\begin{equation}
\chi_A(X) \ = \ \det(X.I_2 - A) \ = \ X^2 \, - \, (\alpha_{11} + \alpha_{22}) X \,  + \, (\alpha_{11} \alpha_{22} - \alpha_{12}.\alpha_{21}) \ . 
\end{equation}

Thus, the matrix $A$ has two real eigenvalues,
\begin{equation}
\left\{
\begin{array}{llll}
        \xi_1 = \frac{\alpha_{11} + \alpha_{22} - \sqrt{(\alpha_{11} - \alpha_{22})^2 + 4\alpha_{12} \alpha_{21}}}{2} \quad ,
     & \\ \xi_2 = \frac{\alpha_{11} + \alpha_{22} + \sqrt{(\alpha_{11} - \alpha_{22})^2 + 4\alpha_{12} \alpha_{21}}}{2} \quad ,
\end{array}
\right.
\end{equation}
$\xi_1 = \xi_2$ is equivalent to $\alpha_{11} = \alpha_{22}$, and $\alpha_{12} = 0$ or $\alpha_{21} = 0$. In that particular case, \\
$\xi_1 = \xi_2 = \alpha_{11} < 1$, and both sequences converge according to Proposition \ref{convergence2Dsuitegamma}, i).

Otherwise, $\xi_1 < \xi_2$, and $A$ is diagonalizable. There exists $W \in \text{GL}_2(\R)$ such that \\
$A = W \begin{pmatrix}
\xi_1 & 0  \\
0 & \xi_{2}
\end{pmatrix} W^{-1}$, that is to say, for $k \in \N$,
\begin{equation}
\label{equation Z_n (forme matricielle développée)}
Z_{k} \ = \ \left[W \begin{pmatrix}
\xi_1^k & 0  \\
0 & \xi_{2}^k
\end{pmatrix} W^{-1}\right] Z_0 \ + \ \left[W \begin{pmatrix}
\frac{1 - \xi_1^k}{1 - \xi_1} & 0  \\
0 & \frac{1 - \xi_2^k}{1 - \xi_2}
\end{pmatrix} W^{-1}\right] L \quad .
\end{equation}
If $\xi_1 = 1$, or $\xi_2 = 1$, we have to replace
$\frac{1 - \xi_1^k}{1 - \xi_1}$ or $\frac{1 - \xi_2^k}{1 - \xi_2}$ by $k$.

As a summary, here,
$(A^k)_{k \in \N} \  \text{converges (towards 0)} \ \, \iff \ \, |\xi_1| < 1 \ \text{and} \ |\xi_2| < 1 \ ,$ and
$(A^k)_{k \in \N} \ \ \text{is bounded} \ \, \iff \ \, |\xi_1| \leq 1 \ \ \text{and} \ \ |\xi_2| \leq 1 \ .$

Regarding these equivalences and Equation \ref{equation Z_n (forme matricielle développée)},
$$ \left(\sum\limits_{i=0}^k A^i \right)_{k \in \N} \ \text{is bounded} \ \ \Rightarrow \ \ 
(A^k)_{k \in \N} \ \ \text{converges towards 0} \ \ \Rightarrow \ \ \left(\sum\limits_{i=0}^k A^i \right)_{k \in \N} \ \text{converges} \ .$$
So, all in all, by also considering the term with $Z_0$, if $(A^k)_{k \in \N}$ converges (towards 0), then $(Z_k)_{k \in \N}$ converges too. 
\smallskip

Now, we look at what happens if $(A^k)_{k \in \N}$ is bounded without converging.

It implies $|\xi_1| \leq 1$, $|\xi_2| \leq 1$, and there exists $i \in \{1,2\}$ such that $|\xi_i| = 1$.

We can show more precisely that $|\xi_1| < 1$, and $\xi_2 = 1$.

Indeed, $\xi_1 + \xi_2 = \alpha_{11} + \alpha_{22} \geq 0$, and 
$\xi_1 \xi_2 = \det(A) = \alpha_{11}  \alpha_{22} - \alpha_{12}  \alpha_{21}$. So $|\xi_1 \xi_2| < 1$.
Thus, if we had $|\xi_1| = 1$, since $\xi_1 < \xi_2$, then $\xi_1 = -1$. But it implies $|\xi_2| < 1$ since $|\xi_1 \xi_2| < 1$, so $\xi_2 + \xi_1 < 0$, which is a contradiction.

Thus, in that case, the sequence $(A^kZ_0)_{k \in \N}$ is bounded, whereas $\left[\left(\sum\limits_{i=0}^k A^i \right) L \right]_{k \in \N}$ diverges. So, all in all, $(Z_k)_{k \in \N}$ diverges. So $(A^k)_{k \in \N}$ cannot be bounded without converging.
\smallskip

Therefore, we get what we wanted, \\
$(Z_k)_{k \in \N} \ \ \text{bounded} \ \ \Rightarrow \ \ (A^k)_{k \in \N} \ \ \text{converges (towards 0)}  \ \ \Rightarrow \ \ 
(Z_k)_{k \in \N} \ \ \text{converges} \ .$
\bigskip

To go further, we may look at more precise conditions on the $\alpha_{ij} , \, 1 \leq i,j \leq 2$ to have the convergence. As a recall, the convergence of $(Z_k)_{k \in \N}$ is equivalent to having $|\xi_1| < 1$ and $|\xi_2| < 1$.

First $|\xi_1| < 1 \iff H^{-}(A) < 2$, where \\
$H^-(A) \, = \, \alpha_{11}^2 + \alpha_{22}^2 + 2 \alpha_{12}\alpha_{21} -(\alpha_{11} +\alpha_{22}) \sqrt{(\alpha_{11} + \alpha_{22})^2 + 4 \alpha_{12} \alpha_{21}} \ .$ \\
Then, $|\xi_2| < 1 \iff H^{+}(A) < 2$, with \\
$H^+(A) \, = \, \alpha_{11}^2 + \alpha_{22}^2 + 2 \alpha_{12} \alpha_{21} + (\alpha_{11} +\alpha_{22}) \sqrt{(\alpha_{11} + \alpha_{22})^2 + 4 \alpha_{12} \alpha_{21}} \ . $ \\
We note that $0 \leq H^-(A) \leq H^+(A)$, so $|\xi_2| < 1 \Rightarrow |\xi_1| < 1$. \\
With some calculus, we deduce that a necessary condition for $H^+(A) < 2$ is \\
$\alpha_{12} \alpha_{21} + \max(\alpha_{11},\alpha_{22})^2 < 1$. 

Moreover, since $|\xi_2| < 1 \Rightarrow |\xi_1| < 1$, we deduce that $|\xi_2|<1$ is \textit{the} necessary and sufficient condition for the convergence of $(\gamma_{1,k})_{k \in \N}$ and $(\gamma_{2,k})_{k \in \N}$.
\medskip

Furthermore, some other sufficient conditions hold by using the relations outlined above, \\
$\xi_1 + \xi_2 = \alpha_{11} + \alpha_{22} \geq 0$, and $|\det A| = |\xi_1 \xi_2| < 1$.

If $\alpha_{11} + \alpha_{22} \geq 1$ and $\alpha_{11} \alpha_{22} \leq \alpha_{12} \alpha_{21}$, then $\xi_1 \leq 0$, and $\xi_2 \geq 1$, and thus, $(Z_k)_{k \in \N}$ diverge. Indeed, $\alpha_{11} \alpha_{22} \leq \alpha_{12} \alpha_{21} \iff \det A = \xi_1 \xi_2 \leq 0$.

Conversely, if $\alpha_{11} + \alpha_{22} < 1$ and $\alpha_{11} \alpha_{22} \geq \alpha_{12} \alpha_{21}$, then $0 \leq \xi_1 \leq \xi_2 < 1$, and $(Z_k)_{k \in \N}$ converge.
\end{proof}

Now, we propose an overview of some conditions related to the convergence of $(\gamma_{1,k})_{k \in \N}$ and $(\gamma_{2,k})_{k \in \N}$. It may be sufficient as well as necessary conditions. However, what corresponds to \textit{the necessary and sufficient condition} is outlined at the end.
As this one is not the most concise, we chose to present some simpler and more understandable conditions before.

\begin{prop}[Conditions for the convergence] ~\\
\label{convergence2Dsuitegamma}
i) If $\alpha_{12} = 0$, or $\alpha_{21}=0$, then the two sequences converge. \\
ii) We assume $\max(\alpha_{11} + \alpha_{21}, \alpha_{12} + \alpha_{22}) < 1$. Then, the sequences converge. \\
iii) Let us assume that $\min(\alpha_{11} + \alpha_{21}, \alpha_{12} + \alpha_{22}) \geq 1$. Then,
$\underset{k \to \infty}{\lim} \gamma_{1,k} = \underset{k \to \infty}{\lim} \gamma_{2,k} = + \infty$. \\
iv) A necessary condition for the convergence is 
$\max(\alpha_{11}, \alpha_{22})^2 + \alpha_{12} \alpha_{21} < 1$. \\
v) a) A sufficient condition for the divergence is having both $\alpha_{11} + \alpha_{22} \geq 1$, and $\alpha_{11} \alpha_{22} \leq \alpha_{12}  \alpha_{21}$. \\ 
b) A sufficient condition for the convergence is having both $\alpha_{11} + \alpha_{22} < 1$, and
$\alpha_{11} \alpha_{22} \geq \alpha_{12} \alpha_{21}$. \\ 
vi) Eventually, the convergence of the sequences \textit{is equivalent} to \\
$\frac{\alpha_{11} + \alpha_{22} + \sqrt{(\alpha_{11} - \alpha_{22})^2 + 4\alpha_{12} \alpha_{21}}}{2} < 1$.
\end{prop}

\begin{proof}
i) If $\alpha_{12} = 0$, $(\gamma_{1,k})_{k \in \N}$ converges by Lemma \ref{lemme gamma_n}. 
It is similar if $\alpha_{21} = 0$.
\medskip

ii) We note $\Phi = \max(\alpha_{11} + \alpha_{21}, \alpha_{12} + \alpha_{22})$, and for $k \in \N$, $\Gamma_k = \gamma_{1,k} + \gamma_{2,k} \geq 0$.

Therefore, 
$\Gamma_{k+1} \leq \lambda_1 + \lambda_2 + \Phi \gamma_{1,k} + \Phi  \gamma_{2,k} = \lambda_1 + \lambda_2 + \Phi \Gamma_k \ .$ We define $(t_k)_{k \in \N}$ by $t_0 = \Gamma_0$ and $t_{k+1} = \lambda_1 + \lambda_2 + \Phi t_k$. We get easily $\Gamma_k \leq t_k$ for $k \in \N$, and since $\Phi < 1$, thanks to Lemma \ref{lemme gamma_n}, $(t_k)_{k \in \N}$ converge, so it is bounded. Hence, $(\Gamma_k)_{k \in \N}$ is bounded, and by positivity, $(\gamma_{1,k})_{k \in \N}$ and $(\gamma_{2,k})_{k \in \N}$ are bounded too, and they converge.
\medskip

iii) With the previous notations, we have this time $\Gamma_{k+1} \geq \lambda_1 + \lambda_2 + \Gamma_k$ for $k \in \N$, which yields $\Gamma_{k+1} - \Gamma_k \geq \lambda_1 + \lambda_2$. So $\underset{k \to \infty}{\lim} \Gamma_k = \infty$. 

We deduce that $\underset{k \to \infty}{\lim} \gamma_{1,k} = \infty$, or $\underset{k \to \infty}{\lim} \gamma_{2,k} = \infty$. 
If we assume that $\underset{n \to \infty}{\lim} \gamma_{1,k} = \infty$, then, since
$\gamma_{2,k+1} = \lambda_2 + \alpha_{21}  \gamma_{1,k} + \alpha_{22}  \gamma_{2,k} \geq \alpha_{21}  \gamma_{1,k} \, , \ $ and $\alpha_{21} >0$
(because $\alpha_{21} + \alpha_{11} \geq 1$ and $\alpha_{11} < 1$), we deduce that $\underset{k \to \infty}{\lim} \gamma_{2,k} = \infty$. 
It would be similar by starting from $\underset{k \to \infty}{\lim} \gamma_{2,k} = \infty$.
\medskip

iv), v), and vi), were already done at the end of the proof of Proposition \ref{caracterisations equivalentes suites gamma(i,k)}.
\end{proof}

\newpage
\section{Ruin probabilities}
\label{section probabilités de ruine}

In this section, we derive some ruin probabilities related to the two models.
After having displayed the results, we discuss the possibility of a comparison between both models, and the sensitivity analysis according to the parameters of each model.

\subsection{The meaning of joint ruin for the carbon model}

As explained in Section \ref{sous-section de présentation du modèle de carbone}, to assess the risks related to the carbon model from Equation \ref{modèle avec BINAR(1) pour carbone}, it makes sense to propose a classical ruin study.
Indeed, it targets situations where an insurance company is responsible for high carbon emissions, or registered too large financial losses, or even both.
Moreover, this should entice the insurers to take care of both non-financial and financial risks jointly (rather than giving priority to financial constraint and focusing less attention on the carbon footprint).

Building on \cite{bettis2017risk}, who argue that cumulative greenhouse-gas emissions expose society to a level of systemic risk that can be interpreted as exceeding the solvency risk typically managed by insurers, we consider how a similar risk language can be operationalised within insurance practice. In a context of emerging risks and evolving regulation, it is plausible that insurers may increasingly be assessed not only on financial solvency, but also on their contribution to emissions through operational decisions (for example). Accordingly, we consider a scenario in which an insurer faces an explicit carbon-footprint limit, interpreted as a firm-level emissions budget for claims-related activities, motivating the study of carbon “solvency” alongside traditional financial solvency. 
\medskip

As a recall, about the meaning of the parameters, we can see $v_2$ as an initial total quantity of carbon which should not be exceeded by a company over a certain duration, offset by an additional quantity $d_2$ allowed per period.
Depending on the context, it could be possible to assume that $d_1$ and $d_2$ are linked through a fixed relation of the type $d_2 = \nu.d_1$, and see therefore $\nu >0$ as a quantity of carbon authorized per financial unit of premium, or, differently but still quite closely, as a quantity of carbon authorized per client insured.

Actually, the main "unofficial" ruin result stated previously is Proposition \ref{prop comparaison conditions de profit net}, which enables, under some assumptions, to deduce the fulfilment of the net profit condition of the carbon model, depending on the one from the classical BINAR(1) model. Besides, Section \ref{section d'analyse des suites gamm(i,k)} provided some characterisations to assess whether this condition is satisfied or not for this classical model.

\subsection{With light-tailed distributions}

In this section, we propose results in case all the distributions of the different types of losses are light-tailed. In particular, for each model, we manage to derive the expression of the corresponding Lundberg equation, and we deduce the existence of an adjustment coefficient.

So as not to merge and confuse the properties related to each model, we first study the model with a classical BINAR(1) process from Equation \ref{équation de définition du modèle en 2D avec BINAR(1)}, then the carbon model from Equation \ref{modèle avec BINAR(1) pour carbone}.

\subsubsection{Results for the classical BINAR(1) model}

We display the moment generating function (m.g.f.) of $S_n$ in the next proposition. Similarly to \cite{ma2015bidimensional} for instance, we express it as a multivariate function, with one variable corresponding to one type of loss. Thus, things are clearer, and it allows us to derive the related Lundberg equation further. Then, we refer to the large deviations theory developed by Glynn and Whitt (in \cite{glynn1994logarithmic}) and its applications to ruin theory (see e.g. \cite{nyrhinen1998rough} and \cite{muller2001asymptotic}) to derive the existence of the adjustment coefficient for our "non-i.i.d." context.

Concerning the subset where it is finite, assuming $M_{X_l}(t) < \infty$ for $l \in \{1,2\}$, \\
${\max} (\alpha_{11}, \alpha_{21})M_{Y_1}(t) < 1$, and
${\max} (\alpha_{12}, \alpha_{22})M_{Y_2}(t) < 1$ should be a sufficient condition. More cautiously, we can replace the two previous conditions by 
$\left(\underset{1 \leq i,j \leq 2}{\min} \alpha_{ij}\right)M_{Y_l}(t) < 1$ for $l \in \{1,2\}$.

\begin{prop}[Moment-generating function in 2D] ~\\
\label{FGM/tf de Laplace en 2D}
Let us assume that the 4 distributions of the losses are light-tailed (respectively those of $X_1$, $X_2$, $Y_1$, and $Y_2$).
Then, the m.g.f. of $S_n$ is, for $(r,s,t,u) \in \R_+^4$ such that $M_{X_1}(r) < \infty$, $M_{X_2}(s) < \infty$, and $\left(\underset{1 \leq i,j \leq 2}{\min} \alpha_{ij}\right)M_{Y_l}(t) < 1$ for $l \in \{1,2\}$,
\begin{equation}
\begin{aligned}
\cM_{S_n}(r,s,t,u) \ &= \ \exp \left[\lambda_1 \left( M_{X_1}(r) M_{Y_1}(t) \sum \limits_{i=0}^{n-1} g_i\left( M_{Y_1}(t), M_{Y_2}(u) \, \right) \right) \right] \\
& \times \, \exp \left[ \lambda_2 \left( M_{X_2}(s) M_{Y_2}(u) \sum \limits_{i=0}^{n-1} h_i \left( M_{Y_1}(t), \,  M_{Y_2}(u) \right) \right) - n(\lambda_1 + \lambda_2) \right] \\
& \times \, \exp \left[ \, \gamma_{1,0} \left(g_n \left( M_{Y_1}(t), \,  M_{Y_2}(u) \right) - 1 \right) \, + \, \gamma_{2,0} (h_n \left( M_{Y_1}(t), \,  M_{Y_2}(u) \right) - 1) \frac{}{} \, \right]  \ ,   
\end{aligned}
\end{equation}
where, for $(t,u) \in \R_+^2$, and $\overline{\alpha_{ij}} = 1 - \alpha_{ij}$,
\begin{equation}
    \left\{
\begin{array}{ll}
        g_0(t,u) = h_0(t,u) = 1 \ ,
     & \\ g_1(t,u) = (\overline{\alpha_{11}} +\alpha_{11} t) (\overline{\alpha_{21}} + \alpha_{21} u) \ ,
     & \\ h_1(t,u) = (\overline{\alpha_{12}} +\alpha_{12} t) (\overline{\alpha_{22}} + \alpha_{22} u) \ .
\end{array}
\right.
\end{equation}
And for $n \geq 1$, 
\begin{equation}
\label{equation relations de récurrence g_n et h_n pour MGF}
\left\{
\begin{array}{ll}
        g_{n+1}(t,u) = g_1(t.g_n(t,u) , \, u.h_n(t,u) ) \ ,
     & \\ h_{n+1}(t,u) = h_1(t.g_n(t,u) , \, u.h_n(t,u) ) \ .
\end{array}
\right.   
\end{equation}
\end{prop}

The proof presents some similarities with the one for Proposition 5 of \cite{Cossette2010DiscreteTimeRM}, or Proposition 3.1 of \cite{ma2015bidimensional}, although we have 4 kinds of losses here instead of 2 or 1 in the models of these papers.

\begin{proof}
We define the following probability-generating function (p.g.f.): \\
$\ P_n(r,s,t,u) := \exp \left( r^{N_{1,1} + \dots N_{1,n}}  s^{N_{2,1} + \dots N_{2,n}}  t^{M_{1,1} + \dots M_{1,n}}  u^{M_{2,1} + \dots M_{2,n}} \right) \, .$ \\
After that, we obtain the m.g.f. simply thanks to the relation: \\
$\cM_{S_n}(r,s,t,u) = P_n \left( M_{X_1}(r), \, M_{X_2}(s), \, M_{Y_1}(t) , \, M_{Y_2}(u) \right) \ .$

It is possible to compute $P_n$ by induction for $n \in \N^*$. 
However, it would be quite heavy to write,  and instead we give some computations for small values of $n$ to give an idea of how it works in general.
\smallskip

For $n=1$, we have 
$P_1(r,s,t,u) = \E \left( r^{N_{1,1}} s^{N_{2,1}} t^{M_{1,1}} u^{M_{2,1}} \right)$. \\
So $P_1(r,s,t,u) = \E \left( (rt)^{N_{1,1}} (su)^{N_{2,1}} \left(t^{\alpha_{11} \circ M_{1,0}} u^{\alpha_{21} \circ M_{1,0}}\right)  \left(t^{\alpha_{12} \circ M_{2,0}}.u^{\alpha_{22} \circ M_{2,0}}\right) \right)$.
By independence of $N_{1,1}$, $N_{2,1}$, $M_{1,1}$, and $M_{2,1}$,
$$P_1(r,s,t,u) = \E \left( (rt)^{N_{1,1}} \right) \E\left((su)^{N_{2,1}}\right) \E \left(t^{\alpha_{11} \circ M_{1,0}} u^{\alpha_{21} \circ M_{1,0}}\right) \E \left(t^{\alpha_{12} \circ M_{2,0}} u^{\alpha_{22} \circ M_{2,0}}\right)  \ .$$
Since $N_{1,1} \sim \cP(\lambda_1)$, $\E \left( (rt)^{N_{1,1}} \right) = \exp [\lambda_1(rt - 1)]$. \\
Similarly, $\E\left((su)^{N_{2,1}}\right) = \exp [\lambda_2(su - 1)]$. \\
Then, we can write:
$
\E \left(t^{\alpha_{11} \circ M_{1,0}}  u^{\alpha_{21} \circ M_{1,0}} 
\right) = \E \left( \E \left(t^{\alpha_{11} \circ M_{1,0}}  u^{\alpha_{21} \circ M_{1,0}} | M_{1,0} \right) \right) \, .
$ \\
Hence, due to the conditional independence of $\alpha_{11} \circ M_{1,0}$ and $\alpha_{21} \circ M_{1,0}$:
$$
\E \left(t^{\alpha_{11} \circ M_{1,0}}  u^{\alpha_{21} \circ M_{1,0}} 
\right) = \E \left( \E \left(t^{\alpha_{11} \circ M_{1,0}} | M_{1,0} \right) \E\left( u^{\alpha_{21} \circ M_{1,0}} | M_{1,0} \right) \right) \ .
$$
Which yields:
$$
\E \left(t^{\alpha_{11} \circ M_{1,0}} u^{\alpha_{21} \circ M_{1,0}} \right) \, =  \E \left( (1 - \alpha_{11} + \alpha_{11} t)^{M_{1,0}}  (1 - \alpha_{21} + \alpha_{21} u)^{M_{1,0}} \right) \, =  \exp [\gamma_{1,0} (g_1(t,u) - 1)] \ .
$$
The calculus is similar for $\E \left(t^{\alpha_{12} \circ M_{2,0}} u^{\alpha_{22} \circ M_{2,0}}\right)$.
\medskip

Similarly, for $P_2$, the factor depending on $N_{1,1}$ is:
$$\E \left( r^{N_{1,1}}  t^{N_{1,1}}  t^{\alpha_{11} \circ N_{1,1}}  u^{\alpha_{21} \circ N_{1,1}}  \right) = \E \left( \E \left( (rt)^{N_{1,1}}  t^{\alpha_{11} \circ N_{1,1}}  u^{\alpha_{21} \circ N_{1,1}} | N_{1,1} \right) \right) \ .$$
Therefore,
$\ \E \left( r^{N_{1,1}}  t^{N_{1,1}}  t^{\alpha_{11} \circ N_{1,1}}  u^{\alpha_{21} \circ N_{1,1}} 
\right)= \E \left( (rt)^{N_{1,1}} \E \left(  t^{\alpha_{11} \circ N_{1,1}}  u^{\alpha_{21} \circ N_{1,1}} | N_{1,1} \right) \right) \ .$ \\
Eventually, 
$\E \left( r^{N_{1,1}}  t^{N_{1,1}}  t^{\alpha_{11} \circ N_{1,1}}  u^{\alpha_{21} \circ N_{1,1}} 
\right) = \exp[  \lambda_1   \left( rt \, g_1(t,u) -1 \right) ] \ .$
\medskip

At a higher stage, the factor in $P_3$ depending on $N_{1,1}$ is: 
$$
E_3 = \E \left( r^{N_{1,1}} t^{N_{1,1}}  t^{\alpha_{11} \circ N_{1,1}}  t^{\alpha_{11} \circ \alpha_{11} \circ N_{1,1}} 
t^{\alpha_{12} \circ \alpha_{21} \circ N_{1,1}} 
u^{\alpha_{21} \circ N_{1,1}}  
u^{\alpha_{21} \circ \alpha_{11} \circ N_{1,1}} 
u^{\alpha_{22} \circ \alpha_{21} \circ N_{1,1}} \right) \ .$$
Once again,
$ \quad
E_3 = \E \left[ \E  ( r^{N_{1,1}} t^{N_{1,1}}  t^{\alpha_{11} \circ N_{1,1}}  t^{\alpha_{11} \circ \alpha_{11} \circ N_{1,1}} t^{\alpha_{12} \circ \alpha_{21} \circ N_{1,1}}  \right.$
$$ \left. \times \, u^{\alpha_{21} \circ N_{1,1}}  
u^{\alpha_{21} \circ \alpha_{11} \circ N_{1,1}} 
u^{\alpha_{22} \circ \alpha_{21} \circ N_{1,1}}| N_{1,1}, \alpha_{11} \circ N_{1,1}, \alpha_{21} \circ N_{1,1} ) \, \right].
$$
So
$$
E_3 = \E \left[ (rt)^{N_{1,1}}  t^{\alpha_{11} \circ N_{1,1}}  [(1 - \alpha_{11} + \alpha_{11}t)(1 - \alpha_{21} + \alpha_{21}u)]^{ \alpha_{11} \circ N_{1,1}} \right.  $$ 
$$ \left. \times \, u^{\alpha_{21} \circ N_{1n}}  
[(1 - \alpha_{12} + \alpha_{12}t)(1 - \alpha_{22} + \alpha_{22}u)]^{\alpha_{21} \circ  N_{1,1}}  \right] \ .
$$
Which corresponds to
$\ E_3 = \E \left[ (rt)^{N_{1,1}}  (t.g_1(t,u))^{\alpha_{11} \circ N_{1,1}}   (u.h_1(t,u))^{\alpha_{21} \circ N_{1,1}} \right] \ .$ \\
It yields eventually:
\begin{equation}
E_3 \ = \ \exp[  \lambda_1  \left[ (rt) \, g_1 (t.g_1(t,u) , u.h_1(t,u) ) -1 \right] ] \ = \ \exp[  \lambda_1  \left[ rt \, g_2(t,u)-1 \right] ]  \ .
\end{equation}
From this type of reasoning, we can infer the expression of $P_n$ for each $n \in \N$.
\end{proof}

With the same assumptions on the losses, it is possible to derive the related Lundberg equation, and deduce the existence of the adjustment coefficient regarding the probability $\pS(u)$.

\begin{cor}[Lundberg equation] ~\\
\label{équation de Lundberg en 2D}
We keep the same assumptions, in particular the 4 distributions of the losses are still light-tailed. We can deduce the Lundberg equation associated with the classical BINAR(1) model.
\end{cor}
\begin{proof}

We notice that $G_n: t \mapsto g_n(M_{Y_1}(t), M_{Y_2}(t))$ and $H_n: t \mapsto h_n(M_{Y_1}(t), M_{Y_2}(t))$ allow to define nondecreasing sequences, in this sense: for all $t \in \R_+$, and $n \in \N$, \\
$G_n(t) \leq G_{n+1}(t)$, and the same thing for $H_n$. 
It may be proved easily by induction.

Therefore, $(G_n)_{n \in \N}$ converges towards a function $g_\infty$. Similarly, $(H_n)_{n \in \N}$ converges towards $h_\infty$. In particular, $G_n(0) = 1$ for each $n$, so $g_\infty(0) = 1$. Similarly, $h_\infty(0) = 1$.

We could work on the whole interval $\R_+$, and $G_n(t)$, $H_n(t)$, $g_\infty(t)$, and $h_\infty(t)$ might be infinite. But from now, we are going to restrict to the maximal interval $I$ such that \\
$g_\infty(t) < \infty$ and $g_\infty(t) < \infty$ for $t \in I$.

It follows the Lundberg equation associated with $t \in I$ is 
\begin{equation}
\label{expression de l'équation de Lundberg en 2D}
(c_1 + c_2)t \  = \ \lambda_1 \, \left( M_{X_1}(t) M_{Y_1}(t)  g_\infty(t) -1 \right) \ + \ \lambda_2 \, \left( M_{X_2}(t) M_{Y_2}(t) h_\infty(t) - 1 \right) \ .
\end{equation}
By writing the relations from Equation \ref{equation relations de récurrence g_n et h_n pour MGF} for $n \to \infty$, it yields: 
\begin{equation}
\label{equations pour g_infty et h_infty}
\left\{
\begin{array}{ll}
        g_{\infty}(t) = g_1 \left( M_{Y_1}(t) g_\infty(t) , \, M_{Y_2}(t) h_\infty(t) \right) \ ,
     & \\ h_{\infty}(t) = h_1 \left( M_{Y_1}(t) g_\infty(t) , \, M_{Y_2}(t) h_\infty(t) \right) \ .
\end{array}
\right.   
\end{equation}
Then, it is possible to deduce the explicit expressions of $g_\infty$ and $h_\infty$.

For the general case, even if the expressions are quite heavy, it is possible to handle them. For $g_\infty$, it yields:
{\small
\begin{equation}
g_\infty(t) \ = \ \frac{2 \, \overline{\alpha_{11}} (\overline{\alpha_{21}} + \overline{\alpha_{12}}M_{Y_2}(t)(\alpha_{21} - \alpha_{22}) )}{1 - \alpha_{11} \overline{\alpha_{21}} M_{Y_1}(t) - \alpha_{22} \overline{\alpha_{12}} M_{Y_2}(t) - (\alpha_{21} - \alpha_{22})(\alpha_{11} \overline{\alpha_{12}} +  \overline{\alpha_{11}} \alpha_{12}) M_{Y_1}(t) M_{Y_2}(t) - \sqrt{\Delta(t)}}  \, ,
\end{equation}
}
where $\Delta(t) = b_1(t)^2 - 4 b_2(t) b_0(t)$, 
and 
{\small
\begin{equation}
\label{expressions des fonctions b_i}
\left\{
\begin{array}{ll}
        b_0 (t) \ = \ \overline{\alpha_{11}} (\overline{\alpha_{21}} + \overline{\alpha_{12}} M_{Y_2}(t) (\alpha_{21} - \alpha_{22}) ) \, ,
     & \\ b_1(t) \ = \ - 1 + \alpha_{11} \overline{\alpha_{21}} M_{Y_1}(t) + \alpha_{22} \overline{\alpha_{12}} M_{Y_2}(t) + (\alpha_{21} - \alpha_{22})(\alpha_{11} \overline{\alpha_{12}} +  \overline{\alpha_{11}} \alpha_{12}) M_{Y_1}(t) M_{Y_2}(t) \, ,
     & \\ b_2(t) \ = \ \alpha_{12} M_{Y_1}(t) M_{Y_2}(t) (\alpha_{11} M_{Y_1}(t) (\alpha_{21} - \alpha_{22}) + \alpha_{22}) \ .
\end{array}   
\right.   
\end{equation}
}

By starting from the two relations of Equation \ref{equations pour g_infty et h_infty}, and by rearranging them, we find that $g_\infty$ must satisfy:
$0 \ = \ b_0(t) + b_1(t)g_\infty(t) + b_2(t) (g_\infty(t) )^2 \ $, where $b_0(.)$, $b_1(.)$ and $b_2(.)$ are the same as in Equation \ref{expressions des fonctions b_i}.

Then, since $\Delta(t)$ is nonnegative, the two possible solutions are \\
$G_-(t) = \frac{- b_1(t) - \sqrt{\Delta(t)}}{2 b_2(t)} \, , \,$ and $G_+(t) = \frac{- b_1(t) + \sqrt{\Delta(t)}}{2 b_2(t)} \, $.

To determine which one is the right solution, we can assess for $t=0$, and use the fact that $g_\infty (0) = 1$. We deduce $G_+(0)$ is the right solution when $t=0$, and then, by continuity of $g_\infty(.)$, $G_+(t)$ is the right solution for each $t \in I$.

Eventually, we rearrange the expression of $g_\infty(t) = G_+(t)$ by multiplying the numerator and the denominator by $- b_1(t) - \sqrt{\Delta(t)}$ : 
$ \ g_\infty(t) = \frac{2 b_0(t)}{- b_1(t) - \sqrt{\Delta(t)}} \ .$

We can check that, when $\alpha_{12} = 0$, we find the same expression as in Equation \ref{expressions g_infty et h_infty quand alpha_12 = 0} below (since $b_2(t) = 0$ in particular).
\medskip

All the reasoning would be the same to determine $h_\infty(t)$. Actually, we can check that both expressions are symmetric, due to the symmetries of the problem.
{\small
\begin{equation}
h_\infty(t) \ = \ \frac{2 \, \overline{\alpha_{22}} (\overline{\alpha_{12}} + \overline{\alpha_{21}}M_{Y_1}(t)(\alpha_{12} - \alpha_{11}) )}{1 - \alpha_{11} \overline{\alpha_{21}} M_{Y_1}(t) - \alpha_{22} \overline{\alpha_{12}} M_{Y_2}(t) - (\alpha_{12} - \alpha_{11})(\alpha_{22} \overline{\alpha_{21}} +  \overline{\alpha_{22}} \alpha_{21}) M_{Y_1}(t) M_{Y_2}(t) - \sqrt{W(t)}}  \, ,
\end{equation}
}
where $W(t) = r_1(t)^2 - 4 r_2(t) r_0(t)$, 
and 
{\small
\begin{equation}
\label{expressions des fonctions r_i}
\left\{
\begin{array}{ll}
        r_0 (t) \ = \ \overline{\alpha_{22}} (\overline{\alpha_{12}} + \overline{\alpha_{21}} M_{Y_1}(t) (\alpha_{12} - \alpha_{11}) ) \, ,
     & \\ r_1(t) \ = \ - 1 + \alpha_{11} \overline{\alpha_{21}} M_{Y_1}(t) + \alpha_{22} \overline{\alpha_{12}} M_{Y_2}(t) + (\alpha_{12} - \alpha_{11})(\alpha_{22} \overline{\alpha_{21}} +  \overline{\alpha_{22}} \alpha_{21}) M_{Y_1}(t) M_{Y_2}(t) \, ,
     & \\ r_2(t) \ = \ \alpha_{21} M_{Y_1}(t) M_{Y_2}(t) (\alpha_{22} M_{Y_2}(t) (\alpha_{12} - \alpha_{11}) + \alpha_{11}) \ .
\end{array}   
\right.   
\end{equation}
}
\end{proof}

\begin{rem}
For the particular case where $\alpha_{12} = 0$, the equation regarding $g_\infty$ can be solved easily. Thus, the expressions are much simpler, in particular the square root disappears.
\begin{equation}
\label{expressions g_infty et h_infty quand alpha_12 = 0}
\left\{
\begin{array}{ll}
    g_{\infty}(t) \ = \ \frac{\overline{\alpha_{11}}(\overline{\alpha_{21}} + M_{Y_2}(t) (\alpha_{21} - \alpha_{22}))}{1 - \alpha_{11} \overline{\alpha_{21}} M_{Y_1}(t) - \alpha_{22}  M_{Y_2}(t) - \alpha_{11}(\alpha_{21} - \alpha_{22}) M_{Y_1}(t)M_{Y_2}(t)} \ ,
     & \vspace{0.2 cm} \\  h_\infty(t) \ = \ \frac{\overline{\alpha_{22}}}{1 - \alpha_{22}M_{Y_2}(t)} \ .
\end{array}
\right.   
\end{equation}
Naturally, the expressions would be symmetric by assuming instead that $\alpha_{21}=0$.
\end{rem}

Then, from the expression of the Lundberg equation, we infer the existence of the adjustment coefficient.

\begin{cor}[Existence of the adjustment coefficient] ~\\
\label{corollaire existence du coeff d'ajustement pour le BINAR}
We keep the same assumptions as before. Then, there exists an adjustment coefficient $\kappa$ such that $\pS(u) \, \overset{\log}{\sim} \, e^{-\kappa(u_1 + u_2)} \,$ when
$u = (u_1,u_2) \to \infty$.
\end{cor}

\begin{proof}
As mentioned previously, we refer to the large deviations theory and its application to ruin theory.

Let us define, for $t \in \R_+$, $f_1(t)$ such that the Lundberg equation may be written $f_1(t) = 0$. According to Equation \ref{expression de l'équation de Lundberg en 2D},
\begin{equation}
\label{equation de Lunberg du BINAR avec f_1}
\begin{aligned}
f_1(t) \  &= \ \lambda_1 \, ( M_{X_1}(t) M_{Y_1}(t)  g_\infty(t) - 1) \\
&+ \ \lambda_2 \, ( M_{X_2}(t) M_{Y_2}(t)  h_\infty(t) - 1)  \, - \,  (c_1 + c_2)t \ . \\
\end{aligned}
\end{equation}

Let us justify $f_1$ is convex, and (at least) $\cC^1$ in 0.

A moment generating function is always nondecreasing, positive, convex, and indefinitely differentiable in a neighbourhood of 0. 

Hence, $f_1$ is $\cC^1$ in a neighbourhood of 0, since $g_\infty(.)$ and $h_\infty(.)$ are $\cC^1$ in 0, as compositions of functions which are $\cC^1$, and because their denominators are not null in a neighbourhood of 0.

Let us give some properties of $g_\infty$. For $n \geq 1$, $G_n$ is convex because the product of two positive, nondecreasing and convex functions is convex, and then, as each $G_n$ is nondecreasing, positive and convex on $I$, $g_\infty$ is also nondecreasing, positive and convex.
It is true for $h_\infty$ too, and by the sum of convex functions, we deduce $f_1$ is convex.

Since $g_\infty(0) = 1$ and $h_\infty(0) = 1$, it yields in particular $f_1(0) = 0$. 

Then, by deriving, we get
\begin{equation}
f_1'(0) = \lambda_1 \mu_{X_1} + \lambda_2 \mu_{X_2} + \gamma_{1,\infty}  \mu_{Y_1} + \gamma_{2,\infty}  \mu_{Y_2} \,  - \, (c_1 + c_2) \ .
\end{equation}
To establish that, we might in particular use the fact $g_\infty(0) = 1$, and the relations it implies between $b_0(0)$, $b_1(0)$, and $b_2(0)$ (as they are defined in Equation \ref{expressions des fonctions b_i}).

thanks to the net profit condition for this model (Proposition \ref{CSprofitneten2D}), $f_1'(0) <0$, hence $f_1$ is negative on a neighbourhood of 0. As before, we note $I = [0,x_0[$ the maximal interval where $f_1$ is defined, namely when $g_\infty$ and $h_\infty$ are finite. We have clearly $\lim\limits_{t \to x_0^-} f_1(t) = + \infty$. Therefore, $f_1$ has at least one root on $]0,x_0[$, and the smallest root is the adjustment coefficient $\kappa$ satisfying the asymptotic relation outlined above.
\end{proof}

\begin{rem}
Actually, we have to be precise and distinguish between the possible types of light-tailed distributions. This previous property is always true for distributions such that $M_X(t) < \infty$ for $t \in [0, M[$, where $M \in ]0,\infty]$, but it would not be necessary true if we had $M_X(t) < \infty$ for $t \in [0, M]$ with $0 < M < \infty$. However, these types are rare, so as not to overcharge the corresponding studies, for the rest of the paper we do not take into account this kind of distribution. 
\end{rem}

It would be difficult to compute the precise adjustment coefficient in general, even for the case of 4 exponential distributions. Indeed, we would obtain heavy or hardly intelligible expressions.
Instead, we provide an upper bound for such a case.

\begin{prop}[Upper bound of $\kappa$ with 4 exponential distributions]~\\
\label{encadrement2Dexpquadruple}
We assume $X_1 \sim \operatorname{Exp}(\alpha)$, $X_2 \sim \operatorname{Exp}(\beta)$, $Y_1 \sim \operatorname{Exp}(\delta)$, $Y_2 \sim \operatorname{Exp}(\eta)$, with $\alpha, \beta, \delta, \eta >0 \, .$ 
Therefore, regarding the coefficient $\kappa$ from Corollary \ref{corollaire existence du coeff d'ajustement pour le BINAR},
\begin{equation}
\kappa \ \leq \ \frac{\alpha + \beta + \delta + \eta}{4} - \frac{\lambda_1 \wedge \lambda_2}{c_1 + c_2} \ .
\end{equation}
\end{prop}

\begin{proof}
We define, for $n \in \N$, the sum of $R_{1,n}$ and $R_{2,n}$, where $C = c_1 + c_2$, and $U = u_1 + u_2$: 
\begin{equation}
Q_n \ = \ U \ + \ C n \ - \ \sum\limits_{k=1}^n \left( \sum\limits_{i=1}^{N_{1,k}} X_{1,k,i} \, + \, \sum\limits_{i=1}^{N_{2,k}} X_{2,k,i} \, + \, \sum\limits_{i=1}^{M_{1,k}} Y_{1,k,i} \, + \, \sum\limits_{i=1}^{M_{2,k}} Y_{2,k,i} \right) \ .
\end{equation}

For $k \geq 1$, since $N_{1,k} \leq M_{1,k}$, and $Y_1$ is a positive variable, we have
$$
2 \sum\limits_{i=1}^{N_{1,k}} \min (X_{1,k,i}, Y_{1,k,i})  \ \leq \ \sum\limits_{i=1}^{N_{1,k}} (X_{1,k,i} + Y_{1,k,i}) \ \leq \ \sum\limits_{i=1}^{N_{1,k}} X_{1,k,i} + \sum\limits_{i=1}^{M_{1,k}} Y_{1,k,i} \ .
$$
And the same thing for the second line.
We define a sequence $(V_k)_{k \in \N^*}$ of i.i.d. variables, such that $V_1 \sim \cP(\lambda_1 \wedge \lambda_2)$. We assume $V$ is independent of $N$, $M$, $X$ and $Y$. We also define an i.i.d. sequence $(Z_{k,i})_{(k,i) \in \N^2}$ such that $Z_{1,1} \sim \operatorname{Exp}(\alpha + \beta + \delta + \eta )$.
We remark that \\ $Z_{k,i} \sim \min \left(X_{1,k,i}, Y_{1,k,i} ,  X_{2,k,i}, Y_{2,k,i} \right)$ for every $(k,i) \in \N^2$.
So, according to the stochastic order:
$$4 \sum\limits_{i=1}^{V_k} Z_{k,i} \leq_{st} \ 2 \left(\sum\limits_{i=1}^{N_{1,k}} \min (X_{1,k,i}, Y_{1,k,i}) + \sum\limits_{i=1}^{N_{2,k}} \min (X_{2,k,i}, Y_{2,k,i}) \right) \, .$$

By defining
\begin{equation}
P_n^{min} \ = \ U \ + \ C \, n \ - \ 4 \sum\limits_{k=1}^n \sum\limits_{i=1}^{V_k} \, Z_{k,i} \ , 
\end{equation}
we get, for $n \in \N$, $Q_n \, \leq_{st} \, P_n^{min}$. 

Hence, with an understandable notation, $\psi_{min}(u) \leq \pS(u)$.

By solving the Lundberg equation
$sC = (\lambda_1 \wedge \lambda_2) \left( \frac{\alpha + \beta + \delta + \eta}{\alpha + \beta + \delta + \eta - 4s} - 1 \right) \, , \ $ 
we find
$$
\psi_{min} (u) \sim D_1 \, e^{- \kappa_1  U} \quad \text{with} \ \ \kappa_1 \ = \ \frac{\alpha + \beta + \delta + \eta}{4} \, - \,  \frac{\lambda_1 \wedge \lambda_2}{C} \ .
$$
Moreover, with this framework, we can apply here the Lundberg inequality for discrete-time risk models to deduce that $0 < D_1 \leq 1$.
\end{proof}

It would have been appropriate to have a lower bound for $\kappa$, because here, such a lower bound would correspond to an upper bound of the risk incurred. Actually, because of the dependence between the counting processes $(N_k)_{k \in \N}$ and $(M_k)_{k \in \N}$, it was more difficult to establish.

Since it is not simple to deduce the expression of the adjustment coefficient for such a model, it should be as well difficult for similar multivariate models with crossing dependence between the lines, but also for unidimensional risk models carrying an autoregression of order $p \geq 2$ (an INAR(p) process typically)-in case this coefficient exists.

\subsubsection{Study for the bivariate carbon model}

In case all the distributions of the losses are light-tailed for the bivariate carbon model, we can compute the associated moment-generating function. As for the previous model, we insist on presenting it as a multivariate function, with one variable corresponding to one type of loss.

\begin{prop}[M.g.f. of the carbon model]
\label{prop MGF en 2D du modèle de carbone}
Let us assume $X_1$, $X_2$, $Y_1$ and $Y_2$ are light-tailed. Then, for $(r,s,t,u) \in \R_+^4$ such that $M_{X_1}(r) < \infty$, $M_{X_2}(s) < \infty$, 
and $\theta M_{Y_1}(t)M_{Y_2}(u) < 1$, we have:
\begin{equation}
\begin{aligned}
\cM_{W_n}(r,s,t,u) \ &= \ \exp \left[\lambda \ M_{X_1}(r) M_{X_2}(s) h_{12} (M_{Y_1}(t) ) h_{21} (M_{Y_2}(u))  \frac{}{} \right.  \\
& \times \,  \left. \left( 1  + h_{\delta} (M_{Y_1}(t) )\sum \limits_{i=1}^{n-1} h_{\beta} \left( g_i \left(M_{Y_1}(t), \,  M_{Y_2}(u) \right) \right) \right) - n \lambda \right] \ .   
\end{aligned}
\end{equation}
where, for $t \in \R_+$, $h_{12}(t) = (1 - \omega_{12} + \omega_{12} t)$, $h_{\delta}(t) = (1 - \delta + \delta t)$, and similarly for $h_{21}(.)$, $h_\beta(.)$, and $h_\theta(.)$.
Moreover, the $g_i$ are defined recursively:
\begin{equation}
    \left\{
\begin{array}{ll}
        g_1(t,u) = t u \, , \ \ \text{and, for} \ \, n \geq 1,
     & \\ g_{n+1}(t,u) = t u \,  h_\theta(g_n(t,u)) \quad .
\end{array}
\right.
\end{equation}
\end{prop}

The computations would be similar to the ones made for Proposition \ref{FGM/tf de Laplace en 2D}, but we do not present them to avoid overcharging the paper.
\bigskip

Then, in relation to the concrete application, we choose, contrary to the previous model, to study also each line separately, because it could correspond to cases where the company is financially ruined, or where there has been too much pollution.

Thus, for $n \in \N^*$, we define $W_{1,n} = \sum\limits_{k=1}^n \left(\sum\limits_{i=1}^{N^C_k} X_{1,k,i} +  \sum\limits_{i=1}^{M^C_{1,k}} Y_{1,k,i} \right)$, and \\
$W_{2,n} = \sum\limits_{k=1}^n \left(\sum\limits_{i=1}^{N^C_k} X_{2,k,i} +  \sum\limits_{i=1}^{M^C_{2,k}} Y_{2,k,i} \right)$.

We can deduce the corresponding m.g.f.:

\begin{prop}[M.g.f. of the financial line]
\label{prop MGF branche financière du modèle de carbone}
Let us assume $X_1$ and $Y_1$ are light-tailed. Then, for $(s,t) \in \R_+^2$ such that $M_{X_1}(s) < \infty$, $\theta M_{Y_1}(t) < 1$. Then, with the previous notations,
\begin{equation}
\cM_{W_{1,n}}(s,t) \, = \, \exp \left[\lambda\left(  M_{X_1}(s) h_{12} (M_{Y_1}(t) )
\left( 1  + h_{\delta} (M_{Y_1}(t) )\sum \limits_{i=1}^{n-1} h_{\beta} \left( q_i \left(M_{Y_1}(t) \right) \right)  \right) - n
\right) \right] \, ,
\end{equation}
where here, for $t \in \R_+$, $q_1(t) = t $, and for $n \geq 1$, $q_{n+1}(t) = t \, h_\theta(q_n(t)) \,$. 
\end{prop}

\begin{prop}[M.g.f. of the ecological line]
\label{prop MGF branche écolo du modèle de carbone}
Let us assume $X_2$ and $Y_2$ are light-tailed. Then, for $(s,t) \in \R_+^2$ such that $M_{X_2}(s) < \infty$, $\theta M_{Y_2}(t) < 1$. Then, with the previous notations,
\begin{equation}
\cM_{W_{2,n}}(s,t) \, = \, \exp \left[\lambda\left(  M_{X_2}(s) h_{21} (M_{Y_2}(t) )
\left( 1  + \sum \limits_{i=1}^{n-1} h_{\beta} \left( l_i \left(M_{Y_2}(t) \right) \right) \right) - n
\right) \right] \, ,
\end{equation}
where here, for $t \in \R_+$, $l_1(t) = t$, and for $n \geq 1$, $l_{n+1}(t) = t  \, h_\theta(l_n(t)) \,$. 
\end{prop}
\bigskip

Then, in terms of asymptotic ruin probabilities, we can derive the existence of an adjustment coefficient in all cases.

We first deal with the general case.

\begin{cor}[Adjustment coefficient $\kappa_C$]
\label{Corollaire coeff d'ajustement général modèle du carbone}
With the same assumptions as for Proposition \ref{prop MGF en 2D du modèle de carbone}, we deduce the existence of an adjustment coefficient related to the model from Equation \ref{modèle avec BINAR(1) pour carbone}, i.e. there exists $\kappa_C$ such that $\pSC(u) \overset{\log}{\sim} e^{-\kappa_C u}$.
\end{cor}

\begin{proof}
We apply the same reasoning as for Corollaries \ref{équation de Lundberg en 2D} and \ref{corollaire existence du coeff d'ajustement pour le BINAR}, in particular in terms of convexity.
From the expression in Proposition \ref{prop MGF en 2D du modèle de carbone}, we deduce the related Lundberg equation which may be written $f_2(t) = 0$, where:
\begin{equation}
\label{equation de Lunberg du carbone avec f_2}
\begin{aligned}
f_2(t) \,  &= \, \lambda \left[ M_{X_1}(t) M_{X_2}(t) h_{12} (M_{Y_1}(t) ) h_{21} (M_{Y_2}(t)) h_{\delta} (M_{Y_1}(t)) h_{\beta} \left( g_\infty (t) \right) \, - \, 1 \right]   \\
&\  - \ (c_1 + c_2)t \ , \\
\end{aligned}
\end{equation}
and
\begin{equation}
g_\infty(t) \ = \ \frac{(1 - \theta) M_{Y_1}(t) M_{Y_2}(t) }{1 - \theta M_{Y_1}(t) M_{Y_2}(t)} \ .
\end{equation}
As it is made in Corollary \ref{corollaire existence du coeff d'ajustement pour le BINAR}, we can prove $g_\infty$ is convex and deduce the existence of an adjustment coefficient, written $\kappa_C$.
\end{proof}

As for the classical model, even for the easiest cases, it seems difficult to compute the adjustment coefficients explicitly, but instead, we can propose an upper bound: 

\begin{prop}[Upper bound of $\kappa_C$ with 4 exponential distributions]~\\
We assume $X_1 \sim \operatorname{Exp}(\alpha)$, $X_2 \sim \operatorname{Exp}(\beta)$, $Y_1 \sim \operatorname{Exp}(\delta)$, $Y_2 \sim \operatorname{Exp}(\eta)$, with $\alpha, \beta, \delta, \eta >0 \, .$
Then:
\begin{equation}
\kappa_C \ \leq \ \frac{\alpha + \beta + \delta + \eta}{4} \, - \, \frac{\lambda \, (\omega_{12} \wedge \omega_{21})}{d_1 + d_2} \ .
\end{equation}
\end{prop}

The proof would be similar to that for Proposition \ref{encadrement2Dexpquadruple}.
\medskip

By considering only one of the two lines, it is still possible to derive such a result.

\begin{cor}
With the same assumptions as for Proposition \ref{prop MGF branche financière du modèle de carbone}, we can deduce the existence of an adjustment coefficient for this financial line, i.e. there exists $\kappa_1$ such that
$\psi_1(u_1) \overset{\log}{\sim} e^{-\kappa_1 u_1}$ when $u_1 \to \infty$.
\end{cor}

\begin{cor}
With the same assumptions as for Proposition \ref{prop MGF branche écolo du modèle de carbone}, we can deduce the existence of an adjustment coefficient for this ecological line, i.e. there exists $\kappa_2$ such that $\psi_2(u_2) \overset{\log}{\sim} e^{-\kappa_2 u_2}$ when $u_2 \to \infty$.
\end{cor}

\begin{proof}
By taking the expressions from Propositions \ref{prop MGF branche financière du modèle de carbone} and \ref{prop MGF branche écolo du modèle de carbone}, we can reproduce the reasoning made in Corollary \ref{Corollaire coeff d'ajustement général modèle du carbone}.
For the first line, the Lundberg equation becomes: 
\begin{equation}
c_1 t \ = \ \lambda \left[ M_{X_1}(t)  h_{12} (M_{Y_1}(t) ) h_{\delta} (M_{Y_1}(t)) \left(1 - \beta + \beta \frac{(1 - \theta) M_{Y_1}(t)}{1 - \theta M_{Y_1}(t)  }  \right)  \, - \, 1 \right] \ .
\end{equation}
And for the second line:
\begin{equation}
c_2 t \ = \ \lambda \left[ M_{X_2}(t)  h_{21} (M_{Y_2}(t) ) \left(1 - \beta + \beta \frac{(1 - \theta) M_{Y_2}(t)}{1 - \theta M_{Y_2}(t)  }  \right) \, - \, 1 \right] \ .
\end{equation}
\end{proof}
\smallskip

If we are interested in $\pO$ and $\pA$, which are linked respectively with the ruin of at least one of the lines, or the ruin of both lines, we can deduce as a direct corollary that they also have an exponential decreasing (related to a logarithmic equivalent) when $(u_1,u_2) \to \infty$. 

Indeed, as mentioned in Section \ref{section d'introduction du modèle avec BINAR(1)}, we have
$\pAC(u) \leq \min(\psi_1(u_1), \psi_2(u_2))$, and \\
$\pOC(u) \leq \psi_1(u_1) + \psi_2(u_2)$.

\subsection{With the presence of heavy-tailed distributions}

In case at least one of the loss distributions is heavy-tailed, we can derive other types of results. We focus mainly on the classical BINAR(1) model in this section, before briefly mentioning the comparison with the bivariate carbon model.
\medskip

If one of the heavy tails is asymptotically dominating the others, we can derive asymptotics for finite-time ruin probabilities. Such results are already presented in \cite{minier2026risk} for the one-dimensional version of the BINAR(1) model, and here it would be proved the same way.
It is also possible if this domination is "shared" by more than one of the distributions. The key properties to derive such results are Proposition X.1.9 and Lemma X.2.2 from the book of \cite{asmussen2010ruin}, but here we prefer not to overcharge the paper with this kind of result or proof.
\medskip

Besides, for multivariate models, there exist a class of distributions which generalises the notion of regularly varying distribution, while settling dependence between the claims: the class of \textit{multivariate regular variations (MRV)} distributions.

We choose the definition used in \cite{konstantinides2016asymptotic}. There are other ways to define the MRV class, such as in \cite{maulik2004characterizations}, or \cite{einmahl2021testing}, but they all refer to the same principle and are very similar.

\begin{defi}[Multivariate regular variations]
Let $d \geq 2$ be an integer. The random vector $(X_1,\dots, X_d)$ is said to follow a multivariate regularly varying distribution if there exists a distribution $H \in \mathcal{R}_{-\alpha}$ and a
nondegenerate limit measure $\nu$ on $[0,\infty]^d \, \backslash  \{0\} $ if for every Borel set $B$ which is $\nu$-continuous (i.e. $\nu(\partial B) = 0$), we have:
$$
\frac{1}{\overline{ H} (t)} \p \left( \frac{1}{t} (X_1, \dots, X_d) \in B \right) \, \underset{t \to \infty}{\longrightarrow} \, \nu(B) \ .
$$
We may note $(X_1,\dots, X_d) \in \operatorname{MRV}(\alpha,\nu, H)$, or $\operatorname{MRV}_{- \alpha}(\nu, H)$.
\end{defi}
\smallskip

From this definition, we infer results when the claims follow a joint MRV distribution. In this way, we relax the assumption made on the mutual independence between the sequences of the amounts of the losses, but we still consider these amounts to be i.i.d. as four-dimensional random vectors.

\begin{prop}{(Ultimate ruin probability for MRV distribution)} ~\\
\label{ruine avec MRV en temps infini}
Let us set $A = \{ (x,y,z,w) \in \R^2 \ / \ x + y + z + w > 1 \}$.
We assume there exists a distribution $H \in \cR_{-\alpha}$, with $\alpha >1$, and a nondegenerate measure $\nu$ on $[0,\infty]^d \, \backslash  \{0\} $ such that $\nu(A) > 0$, $A$ is a $\nu$-continuous set, and
$(X_1, X_2, Y_1, Y_2) \in \operatorname{MRV}(\alpha, \nu,H)$. Yet, we still assume the random vectors $(X_{1,k,i}, Y_{1,k,i}, X_{2,k,i}, Y_{2,k,i})_{(k,i) \in \N^2}$ are i.i.d.
Then, there exists $h \in \cR_{-(\alpha -1)}$ such that $h(u_1 + u_2) \ \leq \ \pS(u_1, u_2)$.
\end{prop}

\begin{proof}
We set, for $(k,i) \in \N^2$, $Z_{k,i} = X_{1,k,i} + Y_{1,k,i} + X_{2,k,i} + Y_{2,k,i}$. Thus, $(Z_{k,i})_{(k,i) \in \N^2}$ is a sequence of i.i.d. variables.
We also set the sequence of i.i.d. variables $(L_k)_{k \in \N}$ such that $L_1 \sim \cP(\lambda_0)$, where $\lambda_0 = \min(\lambda_1,\lambda_2)$. We assume that $L$ is independent of the two-dimension sequences $N$, $M$, $X$ and $Y$. Then, we define, for $n \in \N$,
\begin{equation}
R_n^{min} \ = \ (u_1 + u_2) \ + \ (c_1 + c_2) n \  - \ \sum\limits_{k=1}^n \,  \sum\limits_{i=1}^{L_k} Z_{k,i} \ .
\end{equation}
Therefore, $R_{1,n} + R_{2,n} \, \leq_{st} \, R_n^{min}$. We could apply similar inequalities than for Proposition \ref{encadrement2Dexpquadruple}, thanks to the fact that for $k \geq1$, $N_{1,k} \leq M_{1,k}$ and $N_{2,k} \leq M_{2,k}$, and because the sequences $(N_{1,k})_{k \in \N}$ and $(N_{2,k})_{k \in \N}$ are independent.

Let us take $x >0$.
Thanks to the assumptions, we can write
$$\p(X_{1,1,1} + Y_{1,1,1} + X_{2,1,1} + Y_{2,1,1}  > x) = \p \left( \frac{X_{1,1,1} + Y_{1,1,1} + X_{2,1,1} + Y_{2,1,1}}{x} > 1 \right) \ .$$
So, 
$\, \p(X_{1,1,1} + Y_{1,1,1} + X_{2,1,1} + Y_{2,1,1} > x)  = \p \left( \frac{(X_{1,1,1}, Y_{1,1,1}, X_{2,1,1}, Y_{2,1,1})}{x} \in A \right)$,
which yields $\overline{F_Z}(x) = \p(X_{1,1,1} + Y_{1,1,1} + X_{2,1,1} + Y_{2,1,1}  > x) \sim \nu(A)\overline{H}(x)$, when $x \to \infty$, 
Then, since $H \in \cR_{-\alpha}$, we find $F_Z \in \cR_{-\alpha}$.

In particular, $\psi^{min}(x) \, \sim \, \frac{\lambda_0}{c - \lambda_0 \mu_Z} \, \int_{x}^\infty \overline{F_Z}(y) \, dy$, with $\mu_Z = \mu_{X_1} + \mu_{X_2} + \mu_{Y_1} + \mu_{Y_2}$. 

Thanks to Karamata's theorem (see e.g. \cite{bingham1989regular}), there exists a slowly varying function $L$ such that
$\int_x^\infty \overline{F_Z}(y) \, dy \sim \frac{\nu(A) L(x)}{(\alpha -1) x^{\alpha-1}}$.
\bigskip

It yields the result outlined, with $h(x) = \frac{\lambda_0 \nu(A) L(x)}{(\alpha - 1)(c - \lambda_0 \mu_Z)} \, x^{-\alpha + 1}  \,$.
\end{proof}
\medskip

It is also possible to derive a finite-time result.

\begin{prop}{(Finite-time ruin for a MRV distribution)} ~\\
Let us fix $T \in \N^*$. With the same assumptions as before, there exists  $h_T \in \cR_{-\alpha}$ such that $ h_T(u_1 + u_2) \ \leq \ \pSn(u_1, u_2) $.
\end{prop}

\begin{proof}
For $n \in \N$, we keep the same expression for $R^{min}_n$ as previously.
Then, we can operate successive convolutions (see e.g. Proposition X.1.9 of \cite{asmussen2010ruin}), and we obtain the framework outlined above, with $h_T(x) = \lambda_0 \, T \nu(A) \, \overline{H}(x)$ for $x \geq 0$.
\end{proof}

Once again, it is difficult to derive upper bounds of such ruin probabilities because of the dependence assumptions made for this model.

\begin{rem}
We could modify some assumptions, and assume instead that only the distributions of $X_1$ and $X_2$ are ruled by a MRV distribution, which would be namely here a bivariate regular variations (BRV) distribution, and the $Y_1$ and $Y_2$ are light-tailed, or conversely. We could also separate the lines of business, and assume the pair $(X_1, Y_1)$ is BRV, or conversely. Otherwise, we could assume there are two couples ruled by a BRV distribution... There are numerous possibilities, and it allows to derive such asymptotics not only for $\pS$, but also, with corresponding assumptions, for $\pO$, $\pA$, $\psi_1$, or $\psi_2$.
\end{rem}
\medskip

For the bivariate carbon model from Equation \ref{modèle avec BINAR(1) pour carbone}, we could apply the same types of reasoning to deduce similar asymptotic results.

\subsection{Comparison and sensitivity analysis}

Now, it would be of course interesting to draw a precise comparison between the results of both models.
As it can be made in this type of study (in \cite{minier2026risk} for instance), one would like to apply stochastic ordering properties, such as Theorems 1.A.3 and 1.A.4 from \cite{shaked2007stochastic}. 
Unfortunately, it looks difficult because of the dependence structures involved in each of the models (particularly the classical model), even when trying to propose some reasonable additional assumptions. Nevertheless, regarding the comparisons in terms of risk proposed in Section \ref{section de propriétés basiques}, one would like to assert and prove that the ruin probability of the classical model is an upper bound of that of the carbon model.
\bigskip

In terms of sensitivity, let us look at the classical model first. Intuitively, we would like to prove that the ruin probability $\psi(u)$, seen as a function of any of the parameters ($\lambda_1$, $\lambda_2$, $\alpha_{11}$, $\alpha_{12}$, $\alpha_{21}$, and $\alpha_{22}$), is nondecreasing. Here also, because of the dependence assumptions, it is difficult to prove in general (or at least by using "straightforward" stochastic comparisons). Instead, we can at least focus on the case where the four distributions involved are light-tailed. Indeed, $f_1$, as defined in Equation \ref{equation de Lunberg du BINAR avec f_1}, is a nondecreasing function according to each parameter (for $g_\infty$ and $h_\infty$, this is true by passing to the limit, since it is true for $G_n$ and $H_n$ for each $n \in \N$).
Therefore, $\kappa$ is a nonincreasing function, which means the risk is nondecreasing according to each one of the parameters.

For the carbon model, for the same reasons, it is also complicated to establish a general comparison to encompass the impact of each parameter. Yet, when there are only light tails, by looking at $f_2$, we state once again that $\kappa_C$ is a nonincreasing function of each one of the parameters ($\lambda$, $\beta$, $\theta$, $\delta$, $\omega_{12}$ and $\omega_{21}$). The result is similar for $\kappa_1$ and $\kappa_2$, regarding the respective parameters implied in each case.

\section{Scenarios simulations}
\label{section de simulations de scénarios}

\subsection{Data sources and aggregation method}
\label{sous-section Data sources and Aggregation Method}

The \textit{Livre Blanc} indicates that twelve car pieces account for approximately 73\% of all components restored after collisions and provides average carbon emission values for the three types of restoration operations (repair of the piece, or replacement with a new piece or a PIEC). Namely, these 12 pieces are: 
front bumper, rear bumper, fender, headlights, front door, back door, quarter panel (and rear fender), hood, trunk lid (and tail gate), car spoiler, bumper hubcap, and front chassis (and core support). 

However, the booklet does not specify the corresponding prices, nor the frequency related to each piece. Yet, it mentions that for each claim, there is an average number of 1.16 new pieces and 0.22 PIEC which are used to replace the damaged pieces in case the reparation is difficult or too costly, resulting in a ratio of 15.9\% of the pieces replaced by PIEC. However, as mentioned before, the average carbon emission per claim is 36.2 kgCO$_2$ for the new pieces, but only 1.3 kgCO$_2$ for the PIEC, so they account for only 3.5 \% of the carbon emissions.

In comparison, the eight most frequently damaged parts named by the Allianz report are: front doors, rear doors, windscreen, side panel, headlights, front bumper, rear bumper, and hood. Moreover, contrary to the \textit{Livre Blanc}, this report provides the estimated number of times each part was affected across Europe, based on claims data scaled from Germany. Given this higher resolution of part-specific frequency, we adopt these eight parts as the basis for our costs and emissions estimation model.
Actually, the total number of piece replacements considered by this survey is 48.24 million (see Figure 8 of this report). It concerns only the eight pieces mentioned (and not every possible piece), but the fact that it is a quite large number is significant for our estimations.

Yet, we managed to obtain some figures for the most impacted pieces of a car by some partners involved in the drafting of the \textit{Livre Blanc}.
Indeed, following an annual survey, BCA Expertise, and Crédit Agricole Assurances provided proportions for what must be the 6 most impacted pieces from this survey: front doors, rear doors, front bumper, rear bumper, hood, and side panel. We can state these six parts are included in the eight ones from Allianz, and the twelve ones from the \textit{Livre Blanc}.
More precisely, for each part, we have the average frequency of damages caused by collision claims, and if it is damaged, the frequency of a direct repair (without resorting to piece replacement), or the use of a PIEC.

For consistency, we use the average emission values per action type from the \textit{Livre Blanc}, and apply the Allianz frequency data to simulate the average financial cost per claim related to parts, by combining this with estimated unit prices for new and reused components.
\bigskip

To simulate the financial implications of claim scenarios involving different repair strategies, we estimated the unit costs of parts under the two categories (new pieces or PIEC). These values were not available in the Allianz report; thus, we relied on average prices gathered from specialised automotive retail platforms and sustainability literature.

New part prices were derived from common price ranges observed on European automotive parts marketplaces such as Oscaro, Mister-Auto, and Norauto, as well as from data cited in sustainability and insurance studies, such as the Allianz report, or a document from the ADEME (the French Agency for Ecological Transition). These estimates reflect typical retail prices for parts relevant to mid-range electric vehicles, particularly the Volkswagen ID.3 model used in the Allianz study.

For reused parts (PIEC), we applied a standard 50\% reduction to the new part prices, based on documented practices from industry leaders such as Renault’s Re-Factory and Norauto’s circular parts programs. This choice is consistent with published savings ranges for remanufactured components, typically ranging from 30\% to 70\% off the price of a new part. 

For instance, referring to webpages from different French partners, the price of a PIEC is between 50 and 70\% of the price of a new part according to Thelem Assurances, Société Générale Assurances, Generali or the ADEME, between 50 and 60\% according to Euro Assurance, between 40 and 70\% according to AutoConsultant, from 30\% to 70\% according to MAAF.
Besides, in the United Kingdom, this proportion is around 40 \%, according to \cite{gong2024examining}.

Given the lack of standardised prices across different suppliers and models, the 50\% benchmark offers a balanced and conservative approximation.
\smallskip

The resulting cost structure was combined with frequency data from the Allianz report and the French \textit{Livre Blanc} to compute weighted average values for each scenario, enabling a realistic and policy-relevant simulation of claim management decisions.

\subsection{Estimation of part costs}

Before presenting the rest of the methodology and the figures, we insist on the fact that we are willing to provide conservative lower bounds of the financial costs presented in the remainder of the section. Actually, we already reasoned this way when we considered just above that the price of a PIEC is approximately 50 \% of the price of the same pieces sold as a new part.

We chose to think like that because of the quite broad range of prices for spare parts observed on the different websites or sources mentioned above. Moreover, the corresponding prices should depend on the brand or the type of car. Typically, the prices would be higher than these lower bounds for an SUV. Because of these widely different costs, we also decided not to enumerate the different prices for each part considered coming from each source, so as not to overcharge the paper.
\medskip

As a recall, while the \textit{Livre Blanc} indicates that twelve parts account for 73\% of all components restored after collisions, without specifying the specific proportion of each one, the Allianz report provides explicit frequency data for only eight ones, but does not mention their total joint proportion.

Comparing the two lists of parts in Section \ref{sous-section Data sources and Aggregation Method}, we can see that the eight main parts from Allianz are globally included in the twelve ones from the \textit{Livre Blanc}, excepted the windscreen: this can be explained by the fact that the \textit{Livre} treats with a different approach the case of glass breakage (including the windscreen or the car windows), and it is not taken into account in the figures of piece replacement.
To simplify, in the remainder of the paper, we will consider the eight parts from Allianz that are the most commonly affected. It seems a reasonable assumption based on their inclusion and ordering in the dataset, and since they include key external components such as bumpers, doors, headlights, and windshields, but above all, since it corresponds to a high number of pieces replaced (more than 48 million). Moreover, it seems more prudent to focus only on these eight pieces since, for that, we possess precise figures from Allianz in terms of frequency to draw the estimations, whereas for the other types of parts, such information was difficult to find with enough reliability.

To approximate their representativity, we assume these eight parts capture approximately 80\% of the volume within the twelve, which results in an estimated share of:
$0.80 \times 0.73 = 0.584 \, $.

This implies that our per-claim cost estimations based on these top parts represent approximately 58.4\% of the total average material costs related to post-collision operations. This allows us to scale our results accordingly to approximate full claim-level values.
\medskip

Table \ref{claim_costs} presents the estimated average cost per claim of a new part or a PIEC, based on the eight mentioned parts. As suggested, these estimations are partial, but are still quite representative. Actually, two kinds of parts are gathered in the same category: the front doors or the rear doors. 

\begin{table}[H]
\centering
\begin{tabular}{|c|c|c|c|c|c|c|}
\hline
Part & Prop. (\%) & NP (\euro) & NP prop. (\euro) & PIEC (\euro) & PIEC prop. (\euro) \\
\hline
Windscreen              & 24 & 300 & 71.46 & 150 & 35.73 \\
Front or rear doors & 18 & 350 & 62.11 & 175 & 31.05  \\
Front bumper           & 17 & 450 & 76.68 & 225 & 38.34  \\
Rear bumper            & 15 & 400 & 59.54 & 200 & 29.77 \\
Side panel              & 11 & 400 & 44.78 & 200 & 22.39  \\
Headlights              & 11 & 250 & 27.10 & 125 & 13.55  \\
Hood                   & 4 & 500 & 22.39 & 250 & 11.19  \\
\hline
\textbf{Total}         & 100 \%   & --  & 364.04 & --  & 182.02  \\
\hline
\end{tabular}
\caption{\label{claim_costs}Estimated costs per claim for the eight most frequently affected parts (according to the Allianz report).}
\end{table}

The second column of Table \ref{claim_costs} gives the relative proportion of occurrence for each piece considering all the claims registered in the data from the Allianz report (deduced accordingly to the numbers provided).

In the third and fifth columns, we provide the cost of the piece as a new part or a PIEC. Then, in the fourth and sixth columns, we deduce the respective proportional cost per piece for a given claim. It yields, in the last line, the average cost of a piece, depending on whether it is a new piece or not.

Eventually, to obtain the resulting average part of the financial cost of a claim due to piece replacement, one has to multiply the two respective totals by the average number of pieces replaced per collision claim (which is as a recall, 1.38): it yields \euro  502.38 if only new parts are used, and \euro 251.19 when using exclusively parts from the circular economy. As mentioned, these values may be considered as conservative lower-bound estimates for total material costs per collision.
\bigskip

In comparison, the figures from BCA Expertise and Crédit Agricole Assurances are a little different, and we preferred keeping the ones from Allianz to estimate the prices. However, it is still interesting to display them. In the first column of Table \ref{tab chiffres congrès des actuaires}, we show the average frequency of damage caused to a piece when a collision claim happens.
Then, in the second column, we specify the average frequency of a repair that does not require the use of a piece replacement, and in the third column, the average frequency of the use of a PIEC including all types of claims, i.e. whether a piece replacement is made or not. Then, in the fourth column, regarding the operation of replacement, we deduce the proportion of PIEC used for each kind of piece. 

\begin{table}[H]
\centering
\begin{tabular}{|c|c|c|c|c|}
\hline
Part & Frequency (\%) & Repair (\%) & PIEC (\%) & \% of PIEC among piece repl. \\
\hline
Front doors & 18 & 60 & 23 & 58 \\
Rear doors  & 15 & 61 & 22 & 56 \\
Front bumper & 36 & 34 & 6  & 9.1 \\
Rear bumper  & 7 & 57 & 24 & 56 \\
Side panel   & 26 & 41 & 12  & 20 \\
Hood & 8 & 46 & 11 & 20\\
\hline
\end{tabular}
\caption{\label{tab chiffres congrès des actuaires}Proportions related to pieces repair for six parts (from an annual survey by BCA Expertise and Crédit Agricole Assurances).}
\end{table}

Towards the Allianz report, these figures rather confirm we can merge the role of the front doors or the rear doors, because the corresponding proportions are very close.
Globally, these figures fit with the proportion of repair with or without piece replacement mentioned in the \textit{Livre Blanc}: indeed, for each claim, there is in average 1.38 pieces replaced and 1.03 pieces directly repaired. Here, by considering only these six pieces and taking into account the frequencies in the first column, we have 1.1 pieces damaged per claim, with 0.506 pieces directly repaired and 0.594 pieces replaced.

Besides, one thing that clearly differs from Table \ref{claim_costs} is the number of rear bumpers damaged. Otherwise, by considering the corresponding frequencies properly, for the operation of piece replacement, we observe an approximate use of 30 \% of PIEC. It is much more than the global figures of 15.9 \% for all kinds of pieces mentioned above. We can assume it is due to the fact that it is easier to find a PIEC corresponding to one of these six pieces than for any other type of car piece. Moreover, we can observe that the use of PIEC or a direct repair is much less frequent here for the three front pieces which are essential for people's safety (front bumper, side panel, and hood), whereas PIEC are more used than new pieces for parts which seem less important in that sense (front and rear doors, and rear bumper). It seems consistent, and it should mean the insured client would insist on his safety first, before accepting the use of a PIEC as an environmentally-friendly choice.
\bigskip

What is more, an interesting consideration regarding the eight parts presented in Table \ref{claim_costs} and the various operations described in the \textit{Livre Blanc} concerns the conditions under which a towing service or replacement vehicle might be needed. 

Rather than focusing solely on whether a specific part prevents the vehicle from being driven immediately after the accident, it is more relevant to consider how long the vehicle must remain in the repair shop — a factor that depends on part availability and the complexity of repairs, not only on the nature of the damage. For example, even if the vehicle remains technically drivable after a collision, it may still require a courtesy vehicle if repairs span several days due to logistical delays.

In addition, when reused parts are installed, a different type of inconvenience may arise. If the recycled component fails later — for instance, a door no longer opens properly — the insured might need to return to the garage after the initial claim has been closed. This worsens the overall customer experience, regardless of the severity of the original incident, and generates further indirect costs, including assistance costs.  Namely, these subsequent costs correspond to by-claims in the bivariate carbon model from Equation \ref{modèle avec BINAR(1) pour carbone}, as specified in Section \ref{sous-section de présentation du modèle de carbone}.

The \textit{Livre Blanc} provides estimates of the carbon emissions associated with different assistance operations, though not their financial costs. These include 10.3 kgCO$_2$ for towing, 10.7 kgCO$_2$ when a taxi transports the insured from the accident site, and 154.6 kgCO$_2$ when a replacement vehicle is provided. However, these figures stand for a single operation, and the average emissions per claim for assistance are reported as 10.3 kgCO$_2$, suggesting that in most cases, neither taxis nor replacement vehicles are provided.

\subsection{Simulation of strategic claim scenarios}

To evaluate the impact of different strategic choices by insurers, we simulate various scenarios of client acceptance of PIEC. Based on the estimated average cost and carbon emissions per claim using exclusively new parts, or exclusively PIEC, we calculate weighted average values under intermediate acceptance scenarios.

The \textit{Livre Blanc} acknowledges the carbon emission per operation of replacement for one piece is 6.1 kgCO$_2$ for a PIEC, and 31,4 for a new part. Once again, we multiply by the average number of pieces replaced by claim (1.38) to provide the figures outlined below.
\smallskip

By considering the proportion $\alpha \in [0,1]$ of acceptance of reused parts in the processes of piece replacements, we propose the following numeric representation of the bivariate structure, based on the cost computed previously when PIEC are fully accepted (i.e. $\alpha = 1$), or fully rejected ($\alpha = 0)$:

\[
\mathbf{Z}_k \ = \
\begin{pmatrix}
251.19 \\
8.42
\end{pmatrix}
+
(1 - \alpha) \cdot
\begin{pmatrix}
251.19 \\
34.91
\end{pmatrix}
\]

Thus, the first vector reflects the scenario in which only PIEC is used (financial cost of \euro 251.19 and emissions of 3.8 kg CO$_2$ per claim), and the additive component reflects the additional burden of rejecting PIEC. 
\medskip

Table \ref{T:claim_simulations} summarises the simulated values for different acceptance levels.

The line corresponding to $15.9 \%$ refers to the figures given by the \textit{Livre Blanc}, as mentioned in Section \ref{sous-section Data sources and Aggregation Method}. With our simulation, we find 37.77 kgCO$_2$, which corresponds approximately to the value of 37.5 given in the \textit{Livre}. The slight error must be due to the different levels of approximation between our study and this one.

\begin{table}[H]
\centering
\begin{tabular}{|c|c|c|}
\hline
\% of PIEC acceptance & Cost per claim (\euro) & Emission per claim (kgCO$_2$) \\
\hline
 0\%  & 502.38 & 43.32 \\
 10\%  & 477.26 & 39.84 \\
 15.9 \%  & 462.44 & 37,77 \\
 20\% & 452.14 & 36.35 \\
 30\% & 427.02 & 32.86 \\
 40\% & 401.9 & 29.37 \\
 50\% & 376.79 & 25.88 \\
 65\% & 339.11 & 20.64 \\
 80\% & 301.43 & 15.4 \\
100\% & 251.19 & 8.42 \\
\hline
\end{tabular}
\caption{\label{T:claim_simulations}Simulated average financial and environmental impacts per claim under varying levels of PIEC acceptance.}
\end{table}

Of course, the scenario where close to 100\% PIEC are used is unlikely. 
Indeed, the \textit{Livre Blanc} mentions several possible obstacles: the first one is the affordability of suitable pieces (especially those which are very specific). 
The customer might also be reluctant to use a piece that has already served for another car, in particular if the origin of the piece is uncertain or even unspecified.
One could also mention logistical issues if the corresponding pieces are stocked far from the repair garage.

Thus, the role of the insurers should be to convince their clients to privilege reused pieces from the circular economy when it is possible, for the dual financial and environmental benefit. Actually, he could even be an intermediary or an interlocutor between all the stakeholders: the recycler, the repairer and the insured client. On the other hand, the possibility of a direct repair instead of replacing a piece (by a recycled part or not) remains surely the best financial and ecological solution.

\subsection{The case of electric vehicles}
\label{sous-section véhicules elec}

One would like to know what it could be for the specific case of electric vehicles (EVs).

For this type of vehicle, the trade-offs between economic cost and carbon footprint may differ from those observed for internal combustion engine vehicles.

Unfortunately, the \textit{Livre Blanc} \citep{livreblanc} excluded the electric cars from their study due to insufficient data for robust comparisons, and although this example would offer additional insight to our paper, we do not have consistent data to compare.

However, recent evidence from the France Assureurs (the French insurance federation) suggests that, holding generation comparable, EVs and plug-in hybrids tend to be more expensive to indemnify, largely because they are more costly to repair. Based on an analysis conducted on nearly two million vehicles, some figures and key elements are provided in a corresponding press release \citep{FranceAssureurs2025}.

It is acknowledged in particular that in 2050, more than 50\% of the French vehicles could be electric, or at least hybrid, which underlines a growing trend and the necessity to consider the problem of EVs repairs carefully.
Besides, one may assume the accidents are more frequent during the first months for a person who owns an electric vehicle for the first time, because of the time needed to get used to it.

France Assureurs specifies that the average claim indemnification cost for EVs is about 11\% higher across all coverages, reaching 14\% for own damage coverage and 28\% for glass breakage.
These cost differentials are consistent with repair processes and component constraints that are specific to electrified powertrains. 
Indeed, the electronic architecture increases repair complexity and requires additional caution from the repairer, as an impact can produce subtle effects on these systems.
The report highlights the battery system as a key driver: dismantling, diagnostic procedures, and repair remain sophisticated, and only about half of manufacturers currently offer batteries that are genuinely repairable. This matters directly for claims management because reduced repairability raises the probability that a vehicle is deemed economically irreparable as the EV fleet ages, which can translate into higher total loss frequency and higher average indemnification values.

In addition to battery-related constraints, EVs embed advanced sensors and electronic components in parts that are frequently impacted in accidents. France Assureurs notes that windscreens and headlight optics installed on EVs are, on average, 24\% more expensive to replace, even though accident frequency is not higher for this category. From an operational perspective, these features can increase both parts costs and labour requirements (diagnostics, calibration, safety procedures), and they may extend repair lead times. Longer repair durations also tend to increase the number of rental days for replacement vehicles, with a direct effect on claims expenses, and, depending on the rental fleet mix, the carbon emitted through extra vehicle kilometres and associated energy use.

Vehicle mass is another structural factor. France Assureurs reports that EVs are heavier on average than comparable thermal vehicles, by about 41\% on average, and plug-in hybrids can be heavier by up to 80\%. Higher mass mechanically increases impact energy and can amplify the severity of damage, which in turn raises repair costs. This mechanism is especially relevant for modelling because it provides a plausible link between vehicle technology and the severity distribution of claims, and it may also affect the carbon footprint of repairs through the need for more extensive part replacement and more resource-intensive bodywork processes.

Taken together, these elements suggest that, in the EV setting, cost containment and emission reduction may be less naturally aligned than in conventional repairs. While circular repair options (for example, reuse or refurbished components) can reduce material-related emissions, the feasibility of such strategies may be constrained by safety-critical systems, battery architecture, and restricted access to diagnostic and repair information. France Assureurs explicitly proposes policy levers aimed at preserving affordability while supporting the transition, including strengthening battery repairability standards, ensuring competitive access to diagnostic tools and technical data for independent repairers, and encouraging the use of recycled parts. 

In the same direction, it announces the development of a vehicle “repairability index” (to be elaborated by SRA-Sécurité et Réparation Automobiles-, affiliated with France Assureurs), built on measurable technical elements, with the objective of better informing consumers and anticipating insurance cost impacts, but also to make motor insurance still affordable for everyone.

To conclude this section, even if our empirical analysis does not include EVs, this limitation is not merely a data availability issue: it is also an invitation for future work, since the electric vehicle repair ecosystem is evolving quickly and may change the joint dynamics of claim cost and repair-related emissions over time.

\newpage
\section{Conclusion and continuations}

This paper introduces two bivariate risk models, one of which is designed to capture features of motor-claim management when financial outlays and environmental burdens are considered jointly. Our objective is to support carbon-footprint management in service operations by providing a stochastic bivariate framework that represents, in a unified way, monetary claim costs and repair-related carbon outcomes, thereby enabling decision evaluation under dependence and temporal persistence rather than via independent or static emission accounting. In particular, we aimed to observe whether, and under which conditions, carbon footprint and economic goals are aligned. In that perspective, we model potential dependence between monetary claim costs and ecological costs, the latter being proxied by claim-related carbon emissions, and we argue that treating these two dimensions separately may overlook relevant interactions for both pricing and operational decision-making.

From a methodological standpoint, we establish several risk-related properties of the proposed models. We derive and discuss quantities such as ruin-related measures, covariances, and asymptotical results, and we complement the theoretical developments with numerical illustrations that help compare behaviours across specifications. These elements provide a structured basis to characterise joint risk and to interpret how dependence and persistence may affect both financial and environmental variability in claim dynamics.

On the operational side, we translate part-level replacement choices into an implementable carbon-footprint metric associated with claim management, and we use realistic policy scenarios to stress-test decisions. In particular, varying the acceptance rate of circular-economy reused parts (PIEC) offers a concrete lever through which insurers can evaluate co-benefits: in our illustrations, higher PIEC adoption leads to jointly quantified reductions in expected claim expenditure and CO$_2$-equivalent emissions whenever repair conditions allow. This positions claims management, and specifically repair pathway choices, as an actionable component of decarbonisation strategies in non-life insurance.

Overall, the proposed framework can be read as an insurer-oriented decision tool that quantifies trade-offs and co-benefits between emissions reduction and cost containment. When objectives are aligned, the scenario analysis makes the joint benefit explicit and measurable; when alignment does not hold, the same joint modelling approach allows the trade-off to be stated transparently and translated into operational implications (e.g., decision rules and thresholds for repair pathways) consistent with net-zero and circularity strategies. We also introduce a carbon-footprint ruin perspective, which can be interpreted as assessing exceedance risk relative to an emissions threshold or budget in parallel to financial solvency considerations.

At the same time, we acknowledge the mainly theoretical nature of the present contribution. Its practical relevance would be strengthened by richer and more granular claims data enabling empirical calibration and validation. In particular, estimating the parameters of the proposed bivariate carbon model from real-world portfolios remains a natural continuation, as does a more systematic assessment of the sensitivity of conclusions to emission factors and repair-cost differentials.

Finally, an additional direction concerns demand-side responses: understanding insured clients' acceptance of reused parts and the determinants of such choices is essential to assess the feasibility and effectiveness of circular repair strategies. This behavioural dimension, however, also requires dedicated data and is therefore left for future research.

\newpage

\section{Acknowledgments}

This work benefited from the support of the Research Chair ACTIONS under the aegis of the Risk Foundation, a joint initiative by BNP Paribas Cardif and the Institut des Actuaires.

This work was also conducted within the Research Chair DIALog under the aegis of the Risk Foundation, an initiative by CNP Assurances.
\bigskip

The authors declare no conflicts of interest.
\bigskip

The data were derived from the following resources available in the public domain:
\cite{Allianzrapport2023} and \cite{livreblanc}, documents which are freely available and downloadable online.
\vspace{5 cm}

\appendix

\newpage

\end{document}